\pdfoutput=1 
\documentclass[useAMS,usenatbib]{mn2e}

\usepackage{graphicx}
\usepackage{times}
\usepackage{natbib}
\usepackage{color}
\usepackage{ulem}
\usepackage{lipsum}

\usepackage{hyperref}
\hypersetup{
    colorlinks = true,
    linkbordercolor = {1.0 1.0 1.0},
    allcolors = {cyan},
}

\usepackage[dvipsnames]{xcolor}

\usepackage{amsmath}
\usepackage{amssymb}
\usepackage{float}
\usepackage{todonotes}

\newcommand{\apjs}{ApJS}

\newcommand{\apjl}{ApJL}

\newcommand{\aap}{A \& A}

\defcitealias{dominguezfernandez2020morphology}{Paper~I}

\begin{document}
 
\title[Morphology of radio relics II] {Morphology of radio relics II: Properties of polarised emission}
\author[P. Dom\'inguez-Fern\'andez et al.]{P. Dom\'inguez-Fern\'andez$^{1,2}$\thanks{E-mail: pdominguez@hs.uni-hamburg.de   } , M. Br{\"u}ggen$^1$, F. Vazza$^{3,4,1}$ , M. Hoeft$^5$,
W.~E.~Banda-Barrag\'an$^1$,
\newauthor
K. Rajpurohit$^{3,4}$, D. Wittor$^{1}$, A. Mignone$^{6}$, D. Mukherjee$^{7}$, B. Vaidya$^{8}$  \\
$^{1}$ Hamburger Sternwarte, Universit\"at Hamburg, Gojenbergsweg 112, 21029 Hamburg, Germany\\
$^{2}$Department of Physics, School of Natural Sciences UNIST, Ulsan 44919, Republic of Korea \\
$^3$  Dipartimento di Fisica e Astronomia, Universit\'{a} di Bologna, Via Gobetti 92/3, 40121, Bologna, Italy\\
$^{4}$  Istituto di Radio Astronomia, INAF, Via Gobetti 101, 40121 Bologna, Italy \\
$^{5}$  Th\"uringer Landessternwarte (TLS), Sternwarte 5, 07778 Tautenburg, Germany \\
$^{6}$  Dipartimento di Fisica, Universit\`a di Torino,
  via Pietro Giuria 1, 10125 Torino, Italy \\
$^{7}$ Inter-University Centre for Astronomy and Astrophysics, Post Bag 4, Pune - 411007, India \\
$^{8}$ Discipline of Astronomy, Astrophysics and Space Engineering, Indian Institute of Technology Indore, Khandwa Road, Simrol, Indore 453552, India
}

\date{Received / Accepted}
\maketitle

\begin{abstract}
Radio relics are diffuse radio sources in galaxy clusters that are associated with merger shock waves.
Detailed observations of radio relics in total intensity and in polarisation show complex structures on kiloparsec scales. The relation between the observed features and the underlying morphology of the magnetic field is not clear. Using three-dimensional magneto-hydrodynamical simulations, we study the polarised emission produced by a shock wave that propagates through a turbulent medium that resembles the intracluster medium. We model the polarised synchrotron emission on the basis of diffusive shock acceleration of cosmic-ray electrons. We find that the synchrotron emission produced in a shocked turbulent medium can reproduce some of the observed features in radio relics. Shock compression can give rise to a high polarisation fraction at the shock front and a partial alignment of the polarisation $E$-vectors with the shock normal. Our work confirms that radio relics can also be formed in an environment with a tangled magnetic field. We also  discuss the effect of Faraday Rotation intrinsic to the source, and how our results depend on the angular resolution of observations.

\end{abstract}

\label{firstpage} 

\section{Introduction}
\label{sec:intro}

Radio observations of galaxy clusters reveal Mpc-scale
diffuse emission in the intracluster medium (ICM). \textit{Radio relics} are located at the cluster periphery, with elongated shapes and typically large degrees of polarisation (see \citealt{2012SSRv..166..187B} and \citealt{2019SSRv..215...16V} for reviews). Recent high-resolution radio observations have shown that radio relics have a complex morphology on smaller scales, consisting of filaments, bristles and other substructures that cannot be classified in a single group \citep[e.g.][]{2014ApJ...794...24O,2017ApJ...835..197V,2018ApJ...865...24D,2018ApJ...852...65R,2020A&A...636A..30R,2020ApJ...897...93B,rajpurohit2020understanding}.

Radio relics are believed to trace shock waves generated during mergers of galaxy clusters (see \citealt{1998astro.ph..5367E,2007MNRAS.375...77H,2011JApA...32..505V,2017MNRAS.470..240N, 2020MNRAS.493.2306B} or \citealt{2019SSRv..215...27B} for a review). For those relics that do not resemble merger shocks, it has been suggested that they have their origin in bubbles of relativistic electrons injected by jets from active galaxy nuclei (AGN) that will be re-energised upon a shock passage \citep[e.g.][]{zuhone2020mergerdriven}. One plausible mechanism that accelerates synchrotron-emitting cosmic-ray electrons (CRe) in shocks is the \textit{diffusive shock acceleration} (DSA) \citep[e.g.][]{1987PhR...154....1B,1983RPPh...46..973D}. Nevertheless, only a few radio relics can be explained exclusively by DSA of electrons from the thermal pool \citep[e.g.][]{2020MNRAS.tmpL..75L}. In most cases, the radio power is too large to be explained by DSA of electrons from the thermal pool (see \citealt{2019SSRv..215...16V,2020A&A...634A..64B}).
An alternative model is that a pre-existing population of mildly relativistic CRe ($10 \lesssim \gamma \lesssim 10^4$) exists \citep[e.g.][]{2012ApJ...756...97K,2013MNRAS.435.1061P} that gets re-accelerated by the DSA mechanism. This population of pre-existing CRe could potentially be injected by AGN. 

The observed polarisation fraction in several radio relics is inferred to be locally up to 65\% \citep[e.g.][]{2010Sci...330..347V,2012A&A...546A.124V,2014ApJ...794...24O,2017A&A...600A..18K,2020MNRAS.498.1628L,2020A&A...642L..13R}. 
It is often observed that the polarisation electric field vector ($E$-vector) is aligned with the shock normal, i.e. the polarised magnetic field vector ($B$-vector) is aligned with the shock front \citep[e.g.][]{2009A&A...503..707B,2009A&A...494..429B,2010Sci...330..347V,2017ApJ...845...81P,2017ApJ...838..110G}. Other observations of radio relics show similar trends even though one should note that in some cases the $E$-vectors are not corrected for Faraday rotation \citep[e.g.][]{2012MNRAS.426.1204K,2014MNRAS.444.3130D,2015MNRAS.453.3483D}. A compilation of polarised radio relics can be found in Table 1 of \citet{wittor2019}. The alignment could either be produced by a large-scale uniform or compression of a randomly oriented magnetic field \citep[e.g.][]{1980MNRAS.193..439L,1998astro.ph..5367E}. It is also not clear whether the typically observed mild strength of shocks ($\mathcal{M}_{\mathrm{radio}}$\footnote{ $\mathcal{M}_{\mathrm{radio}}$ is the Mach number inferred from radio observations.} $\sim 1.7$--$4.6$) would be enough to explain this alignment \citep[e.g.][]{2006AJ....131.2900C,2010Sci...330..347V,2012A&A...546A.124V}. Furthermore, the observed degree of alignment in relics might also depend on the viewing angle if the polarisation is caused by compression of a small scale tangled magnetic field. In particular, a relic viewed face-on is expected to have a less coherent alignment than the edge-on view \citep[e.g.][]{2013ApJ...765...21S,wittor2019}. This happens because only the perpendicular components of the magnetic field with respect of the shock's normal are affected by the shock, while the parallel component is conserved. On the other hand, 
the observed degree of alignment and polarisation would not depend on the viewing angle if the observed region of the ICM underlines a large scale homogeneous uniform magnetic field.

The high polarisation fraction, particularly at high radio frequencies, in radio relics makes them ideal objects for studying magnetic field properties of the ICM. However, modelling the observed features of radio relics is challenging from a numerical point of view. On one hand, cosmological simulations lack the resolution to solve the CRe's cooling scales (e.g. \citealt{2013ApJ...765...21S,2015ApJ...812...49H,2017MNRAS.470..240N,wittor2019}). On the other hand, Particle In Cell (PIC) simulations can only tell us about the dynamics at microphysical scales \citep[e.g.][]{2014ApJ...797...47G,2014ApJ...783...91C,2015PhRvL.114h5003P,2014ApJ...794...46C,2019ApJ...883...60R,2019ApJ...876...79K}. Therefore, it is relevant to bridge the gap of spatial scales between PIC and cosmological simulations.
In \citet{dominguezfernandez2020morphology} (hereafter \citetalias{dominguezfernandez2020morphology}), we modelled the synchrotron emission in a small fraction of the ICM by means of a new hybrid particle and fluid framework using the magneto-hydrodynamical (MHD) code PLUTO \citep[][]{pluto1,2018ApJ...865..144V}. This method uses \textit{Lagrangian} particles, each with its individual energy spectrum, embedded in a large-scale MHD simulation.

In \citetalias{dominguezfernandez2020morphology}, we studied a set-up where a shock was launched in an initially turbulent ICM. We studied how the properties of the upstream ICM are related to the synchrotron properties of radio relics assuming DSA. Some of the main results of \citetalias{dominguezfernandez2020morphology} are:

\begin{itemize}
    \item[i)] The existence of complex substructure in the radio maps is a result of turbulence in the ICM.
    \item[ii)] The interaction of a shock with a turbulent medium can reproduce the discrepancy between Mach numbers as inferred from X-ray and radio emission. We found that one cause of this discrepancy is the stronger dependence of the synchrotron emission on the compression in the shock and therefore, the higher Mach numbers in the tail of the Mach number distribution. We also found that the amplitude of the magnetic field fluctuations decreases more slowly than the density and temperature fluctuations, which adds to the observed Mach number discrepancy.
    \item[iii)] The magnetic power spectrum of a decaying turbulent medium remains largely unaffected at $\lesssim 10$ kpc scales by shocks with typical sonic Mach numbers of $\mathcal{M}=2$--3. Independently of the initial type of turbulence, we found that shock compression and propagation shift the magnetic power spectra towards higher wave numbers (smaller scales) initially, and afterwards it generates post-shock turbulence leading to a shift of the turbulent power towards smaller wave numbers (larger scales).
    \item[iv)] $\mathcal{M}=2$ shocks are not strong enough to modify the initial pre-shock magnetic field. This implies having an unrealistically high acceleration efficiency.
\end{itemize}

In this work, we extend our study of \citetalias{dominguezfernandez2020morphology} by including the modelling of polarised emission. Our aim is to study how the structure of the radio emission that is now observed in various radio relics would be observed in polarisation. A recent high-resolution and high-sensitivity polarimetry study of the cluster CIZA J2242.8+5301 \citep[e.g.][]{2021arXiv210206631D} indicates that the polarisation emission can decrease towards the downstream. We expect that other highly resolved relics such as the Toothbrush relic in the galaxy cluster 1RXS J0603+4214 \citep[e.g.][]{2018ApJ...852...65R} or the MACS J0717.5+3745 relic \citep[e.g.][]{2009A&A...503..707B,rajpurohit2020understanding}, would give us also more information about the downstream characteristics of relics when observed in polarisation. 

We consider the same set-up as in \citetalias{dominguezfernandez2020morphology}, i.e., we drive a shock through a turbulent medium which is representative of a small region of the ICM. We then assume that CR-electrons are injected instantly at the shock discontinuity and acquire an initial energy distribution based on DSA. The paper is structured as follows: in Sec.~ \ref{section:num_set-up2}, we describe our numerical set-up and initial conditions. In Sec.~\ref{sec:pol_methods}, we include a description of the polarised emissivity and explain how we obtain the Stokes U and Q  parameters maps. Sec.~\ref{sec:results2} shows our results for a uniform and a turbulent ICM and finally, we summarise in Sec.~\ref{sec:conclusions2}. Throughout this paper, we assume a $\Lambda$CDM cosmology \citep[e.g.][]{planck2018}.

\section{Numerical set-up}\label{section:num_set-up2}

\begin{figure}
    \centering
    \includegraphics[width=\columnwidth]{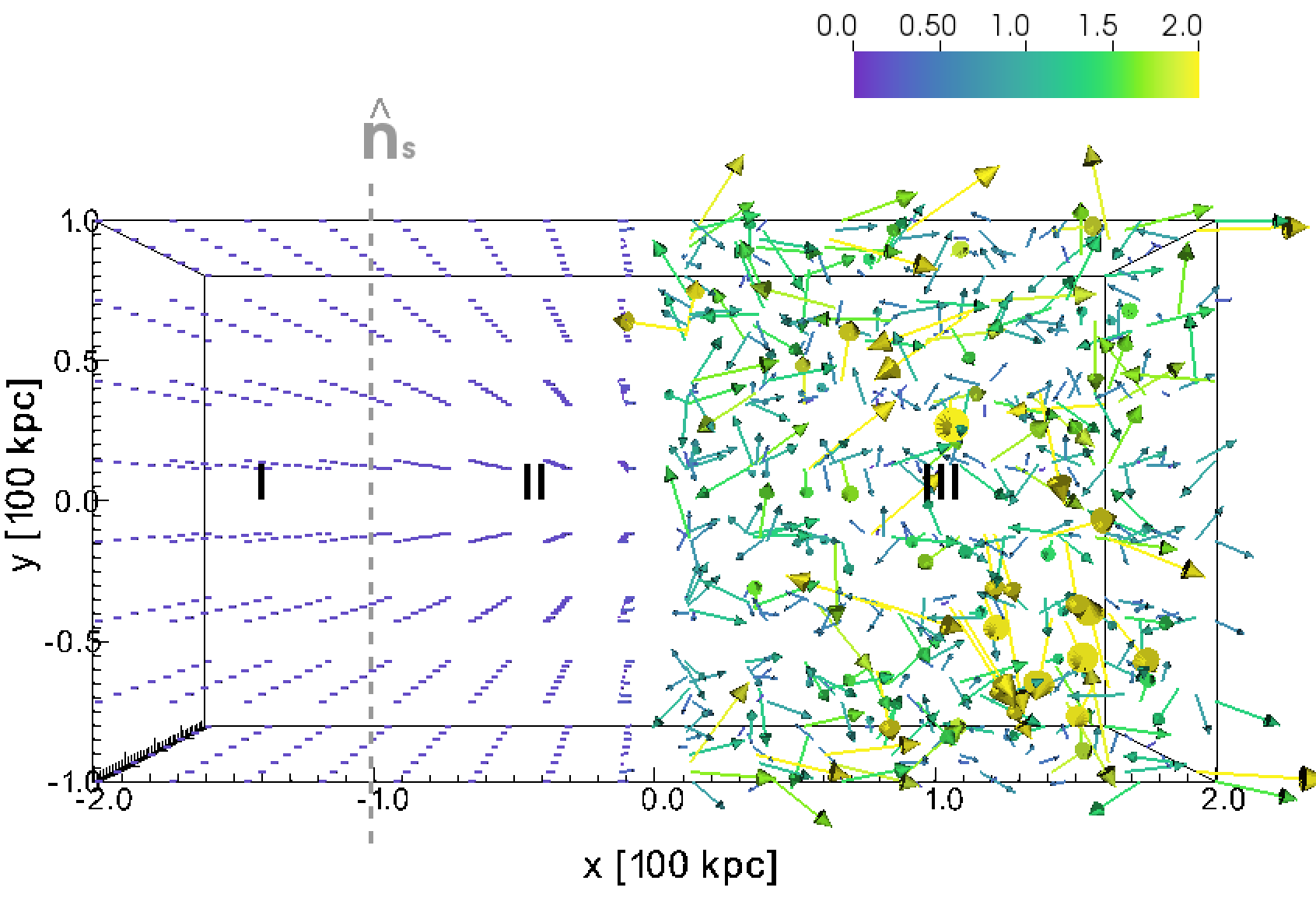}
    \caption{Initial magnetic field configuration in the PLUTO code. The vectors are coloured according to the magnitude of the magnetic field as shown in the upper colourbar in units of $\mu$G. I denotes the post-shock region, II denotes the uniform pre-shock region and III the turbulent pre-shock region. The left side is a uniform medium with a $B_x$ component matching the mean value of the $B_x$ of the turbulent medium. We have one Lagrangian particle per cell placed in the whole regions II and III. The dashed gray line denotes the location of the shock discontinuity.}
    \label{fig:init}
\end{figure}

The basic set-up of our simulation is the same as in \citetalias{dominguezfernandez2020morphology}. Our computational domain is a rectangular box (400 kpc $\times$ 200 kpc $\times$ 200 kpc with $256 \times 128 \times 128$ cells, respectively), where $x \in$ [-200,200] kpc, $y \in$ [-100,100] kpc, and $z \in$ [-100,100] kpc. The right half of the domain is filled with a turbulent medium, representing a realistic ICM. The left half contains a uniform pre- and post-shock medium, between which a shock is launched. We define a shock discontinuity at $x=-100$ kpc (see Fig.~\ref{fig:init} for the initial configuration of the magnetic field). This defines three regions in our simulation box: a post-shock uniform region (I), a pre-shock uniform region (II) and a pre-shock turbulent region (III) (see Fig.~\ref{fig:init} and Sec.~2.2 in \citetalias{dominguezfernandez2020morphology}).

\begin{figure}
    \centering
    \includegraphics[width=0.9\columnwidth]{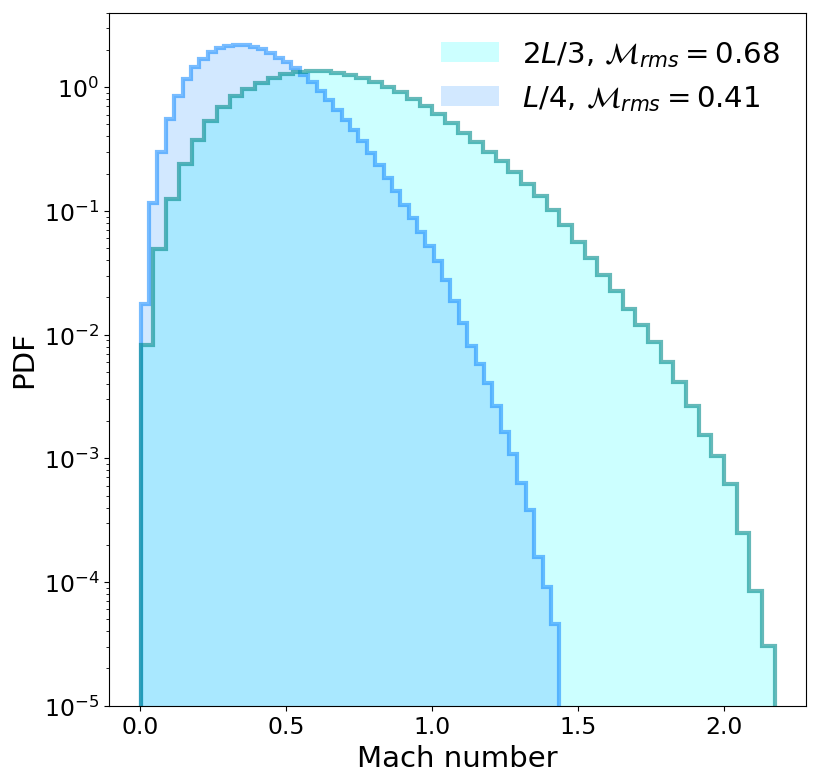}
    \includegraphics[width=0.9\columnwidth]{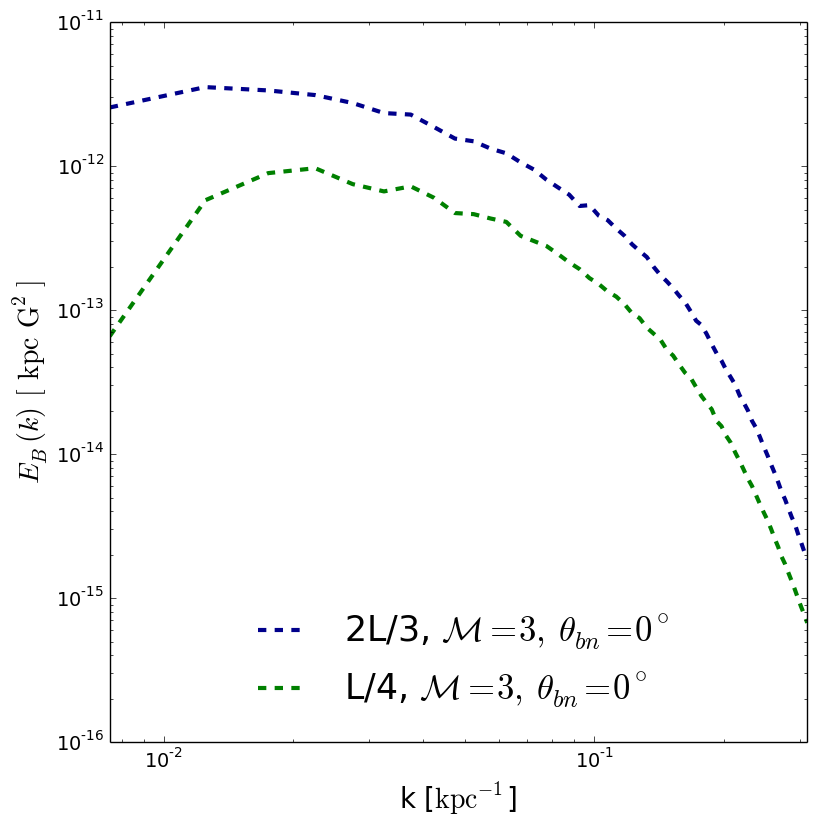}
    \caption{\textit{Upper panel}: Initial distribution of Mach numbers in the $2L/3$ and $L/4$ turbulent media. \textit{Lower panel}: Initial magnetic power spectra of the $2L/3$ and $L/4$ turbulent media.}
    \label{fig:mach_flash}
\end{figure}

The turbulent ICM initial conditions for region (III) were produced using the MHD FLASH code, version 4.6.1 \citep{2000ApJS..131..273F, 2002ApJS..143..201C}. Each initial condition is self-consistently turbulent, i.e. the density, velocity, temperature, and magnetic field are turbulent fields. We produced two different turbulent media whose main characteristics are summarised as\footnote{For all specifics regarding these FLASH simulations we refer the reader to Sec.~2.1 in \citetalias{dominguezfernandez2020morphology}.}:
\begin{itemize}
    \item[i)] $2L/3$ \textit{turbulent medium}: Power peaks in 2/3 of the 200 kpc box, representing an injection scale of 133 kpc. Plasma-beta of $\sim 110$ and an rms sonic Mach number of $\mathcal{M}_{\rm rms} \sim 0.68$ (see Fig.~\ref{fig:mach_flash}). This medium has a mean density of $0.8 \times 10^{-3} \, \mathrm{cm}^{-3}$ and a mean magnetic field of 1 $\mu$G.
    
    \item[ii)] $L/4$ \textit{turbulent medium}: Power peaks
in 1/4 of the 200 kpc box, representing an injection scale of 50 kpc. Plasma-$\beta$ of $\sim 110$ and an rms sonic Mach number of $\mathcal{M}_{\rm rms} \sim 0.41$ (see Fig.~\ref{fig:mach_flash}). This medium has a mean density of $0.8 \times 10^{-3} \, \mathrm{cm}^{-3}$ and a mean magnetic field of 1.5 $\mu$G,
\end{itemize}
where the plasma-$\beta$ is defined as the ratio of the plasma pressure to the magnetic pressure, $\beta=8\pi P/B^2$. The rms Mach number $\mathcal{M}_{\rm rms}$ is an integral quantity defined as $(\int v^2 dV/ \int c_s^2 dV)^{1/2}$, where $v$ is the velocity field, $c_s$ is the sound speed and $V$ is the volume of the computational domain. In the lower panel of Fig.~\ref{fig:mach_flash}, we show the magnetic power spectra of both initial conditions in which the injection scale or peak spectra is better visualised.

We used the ideal MHD code PLUTO \citep{pluto1} to evolve this system in the presence of shock acceleration and then compute the polarised emission. We assumed an ideal equation of state (EOS), that is an adiabatic index $\gamma_{0} = 5/3$. The initial boundary conditions of the computational domain are \textit{outflow} in $x$ (zero gradient across the boundary) and \textit{periodic} in $y$ and $z$. We used a piece-wise parabolic method (PPM) for the spatial integration, whereas a $2^{nd}$ oder TVD Runge-Kutta method for the time stepping with a Courant-Friedichs-Lewy (CFL) condition of $0.3$. The Riemann solver for the flux computation that we used is the Harten-Lax-van Leer-Discontinuities (HLLD) solver \citep[see][]{2005JCoPh.208..315M}. We control the $\nabla \cdot \mathbf{B} = 0$ condition with the hyperbolic divergence cleaning technique where the induction equation is coupled to a generalised Lagrange multiplier (GLM) \citep[e.g.][]{2002JCoPh.175..645D}.

 We performed simulations with both turbulent media from \citetalias{dominguezfernandez2020morphology}, namely the $2L/3$ and $L/4$ cases. The initial conditions for the density, pressure, and velocity in region II (\textit{pre-shock} uniform region at [-100,0] kpc) 
 are set to the mean value of the corresponding turbulent fields. In the case of the magnetic field in region II, we set it to be the mean value of the $B_x$ component of the turbulent medium. The initial conditions for region I (\textit{post-shock} region) are selected according to the MHD Rankine-Hugoniot conditions \citep[e.g.][]{Landau1987Fluid}. We set up an initial shock with sonic Mach number $\mathcal{M}$ and study the polarisation as observed from different lines of sight (LOS) and frequencies. Finally, we fill the computational domain from the shock discontinuity up to the right side of the box with one \textit{Lagrangian} particle per cell. This gives us a total number of 3,145,728 Lagrangian particles for each run. Each particle evolves according to the following prescription:
 \begin{table}
\centering
\begin{tabular}{ccccc}
    &&& \\ \hline
      Uniform medium & $\mathcal{M}$ &$\theta_{bn}[^{\circ}]$ &$\rho_{\mathrm{II}} \, [10^{-27}\mathrm{g/cm}^3] $ & 
       $B_{\mathrm{II}} \, [\mu G]$ \\ \hline
       
       1 & 3.0  & 0 & 1.34 & 0.4 \\
       2 & 3.0  & 90 & 1.34 & 0.4 \\
       
    &&& \\ \hline
      Turbulent medium & $\mathcal{M}$ &$\theta_{bn}[^{\circ}]$ &$\rho_{\mathrm{II}} \, [10^{-27}\mathrm{g/cm}^3] $ & 
       $B_{\mathrm{II}} \, [\mu G]$ \\ \hline
      $2L/3$ & 3.0  & 0 & 1.34 & 0.4 \\ 
      $L/4$ & 3.0 & 0 & 1.34 & 0.4 \\ 
      \hline
\end{tabular}
\caption{Initial conditions of region II: uniform pre-shock region ([-100,0] kpc for the turbulent medium and [-100,200] for the uniform medium). The initial conditions for the left side of the shock (region I) depend on the pre-shock conditions (region II) and the initial Mach number of the shock $\mathcal{M}$ through the Rankine-Hugoniot jump conditions. $L$ denotes the length of the turbulent region, i.e. 200 kpc. $\theta_{bn}$ is the angle of the upstream magnetic field with respect to the normal of the shock. Note that the magnetic field in region II has only an x-component, $B_{x,\mathrm{II}}$,  in the turbulent media. The two runs with turbulent media correspond to the IDs k1p5\_M3\_parallel and k4\_M3\_parallel in Table 1 in \citetalias{dominguezfernandez2020morphology}.}
\label{table:init2}
\end{table}
 \begin{itemize}
     \item[i)] \textit{Activation}: each Lagrangian particle is activated once it is located in a shocked cell
     (see Sec.~3.1 and Appendix B in \citetalias{dominguezfernandez2020morphology} for a description of the activation and the shock finder algorithm). At that moment, the particle acquires an energy spectrum according to DSA theory (see Sec.~3.1 in \citetalias{dominguezfernandez2020morphology}). We assumed a fixed acceleration efficiency of $\eta=10^{-3}$ and fixed energy limits of $\gamma_{\rm min} = 1$ and $\gamma_{\rm max} = 10^5$ (energy high enough to model observations at 6.5 GHz in a $\sim 1 \, \mu$G magnetic field region).
     The final synchrotron emission can be re-scaled to other values of $\eta$ since in our case the energy limits remain constant and the accelerated CRe do not have any feedback on the shock evolution. We defer to future work a more extensive study on the role of different model choices for $\eta$ and $\gamma_{min}$ (which are poorly constrained in the DSA theory of cosmic ray electron acceleration; see \citealt{2001RPPh...64..429M} and references therein for a general review on DSA at collisionless shocks and its challenges).

     \item[ii)] \textit{Evolution}: The energy spectra of the particles evolve according to a cosmic-ray transport equation involving adiabatic, synchrotron, and inverse Compton losses \citep[see][]{2018ApJ...865..144V}.
 \end{itemize}
 
 All the parameters of our runs are summarised in Table~\ref{table:init2}.
 Shocks can be classified as quasi-parallel and quasi-perpendicular if $\theta_{bn} \leq 45^{\circ}$ or $\theta_{bn} > 45^{\circ}$, respectively, where $\theta_{bn}$ refers to the angle of the upstream magnetic field with respect to the shock normal. In this work, we consider for simplicity only two limits, i.e. $\theta_{bn} = 0^{\circ}$ and $\theta_{bn} = 90^{\circ}$ for the initially uniform medium and $\theta_{bn} = 0^{\circ}$ for the initially turbulent medium. In the case of the turbulent medium it should be stressed that the initial $\theta_{bn} = 0^{\circ}$ refers only to the initial angle of the average upstream magnetic field defined in region II with respect to the shock normal. Note that the magnetic field in region II only has an $x$-component and its value is defined such that it matches the mean value of the $B_x$ component of the turbulent medium (region III). As the shock propagates through region III, $\theta_{bn}$ changes from cell to cell depending on the local magnetic topology. Nevertheless, throughout the paper we will use the notation $\theta_{bn} = 0^{\circ}$ as a reference to the initial configuration. We refer the reader to \citetalias{dominguezfernandez2020morphology} for a more detailed description of the initial set-up.
 
 For each configuration in Table~\ref{table:init2}, we produced polarisation maps at three frequencies: 150 MHz, 1.5 GHz, and 6.5 GHz. 
 We restricted ourselves to consider only these configurations (see Table \ref{table:init2}) because in \citetalias{dominguezfernandez2020morphology} we found that $\mathcal{M}=2$ shocks in our set-up are unlikely to reproduce observable radio relics. In \citetalias{dominguezfernandez2020morphology}, we also found differences in the synchrotron emission computed with the two different turbulent media. Here we focus on the correspondent differences produced in the polarised emission. We also leave the cases with $\theta_{bn}=90^{\circ}$
 out of this work as the study of the shock obliquity will be subject of our study in the third paper in this series.

\section{Polarised emission from shocks}\label{sec:pol_methods}

The linearly polarised emission is computed in a similar fashion as the synchrotron emission \citep[see][]{1965ARA&A...3..297G}:
\begin{equation}\label{J_pol2}
     \mathcal{J}_{\rm pol}^{'}(\nu_{\rm obs}',\mathbf{\hat n_{los}'},\mathbf{B}') 
     = \frac{\sqrt{3} e^3}{4\pi m_e c^2} 
     | \mathbf{B}' \times \mathbf{\hat n_{los}'}| \int N(E') G(\xi) \, dE',
\end{equation}
where $\mathbf{B}'$ is the local magnetic field, $\mathbf{\hat n_{los}}'$ is the unit vector in the direction of the LOS in the comoving frame. Each Lagrangian particle (or macro-particle) that represents an ensemble of CR electrons is characterised by an energy distribution function, $N(E)$ (as defined in \citetalias{dominguezfernandez2020morphology}). The macro-particle's energy distribution function at the activation time is
\begin{equation}
\chi(E) = \frac{N(E)}{n_0} = \frac{N_0}{n_0}\, E^{-p},     
\end{equation}
where $p=q-2$ is the energy \textit{injection spectral index}, $q$ is the power--law index of the corresponding particle momentum distribution and related to the shock Mach number $\mathcal{M}$ via the DSA theory, $N_0$ is the normalisation factor, and $n_0$ is the fluid number density (see Sec.~3.1 of \citetalias{dominguezfernandez2020morphology}). 
$N_0$ is assigned according to the kinetic energy contained in the shock. That is, we considered that the total energy per fluid number density is
\begin{equation}
    \int \chi(E)\, E \, dE = \frac{E_{tot}}{n_0},
\end{equation}
where $E_{\rm tot} = \eta \, E_{\rm shock} = \eta \, \frac{1}{2}\rho_{\rm post} \,v_{\rm shock}^2$ and $\eta$ is the acceleration efficiency. From these, we finally obtain the normalisation factor:
\begin{equation}\label{eq:norm}
    N_0 = \left\{
	     \begin{array}{ll}
		 \frac{\eta \, E_{\rm shock}\,(4-q)}{\left[{E_{\rm max}}^{4-q}\, - \, {E_{\rm min}}^{4-q} \right]}       & \mathrm{if\ } q \ne 4 \\
		 & \\
		 \eta \, E_{\rm shock} \log \left( \frac{E_{\rm max}}{E_{\rm min}}\right) & \mathrm{if\ } q=4 \\
		 
	       \end{array}
	     \right.
\end{equation}

Finally, $G(\xi)=\xi K_{2/3}(\xi)$, where $K_{2/3}(\xi)$ is a modified Bessel function and 
\begin{equation}
    \xi = \frac{\nu_{\rm obs}'}{\nu_c^{'}} = \frac{4\pi m_{e}^3c^5 \nu_{\rm obs}'}{3eE'^2| \mathbf{B}' \times \mathbf{\hat n_{los}'}|},
\end{equation}
where $\nu_{\rm obs}'$ and $\nu_c'$ are the observing and critical frequencies, respectively. The reader should note that here the primed quantities refer to the comoving frame, whereas standard notation refers to the observer's frame. Note also that only those particles with pitch angle coinciding with the angle between $\mathbf{B}'$ and $\mathbf{\hat n_{los}'}$ contribute to the emission along the LOS in Eq.~ (\ref{J_pol2}).

The polarised emissivity in Eq.~(\ref{J_pol2}) is measured in the local comoving frame with the emission region. To obtain the emissivity in a fixed observer's frame, we have to apply a transformation:
\begin{equation}
    \mathcal{J}_{\rm pol}(\nu_{\rm obs},\mathbf{\hat n_{los}},\mathbf{B}) = \mathcal{D}^2
    \mathcal{J}_{\rm pol}^{'}(\nu_{\rm obs}',\mathbf{\hat n_{los}'},\mathbf{B}'), 
\end{equation}
where $\mathcal{D}$ is a Doppler factor (see Eq. (21) in \citetalias{dominguezfernandez2020morphology}).
The unit vectors in the direction of the line of sight in the comoving and observer's frame are related via
\begin{equation}\label{eq:nlos_comoving}
    \mathbf{\hat n_{los}'} = \mathcal{D} \left[ 
    \mathbf{\hat n_{los}} + \left( \frac{\gamma^2}{\gamma +1} \mathbf{v} \cdot \mathbf{\hat n_{los}} - \gamma
    \right) \mathbf{v}
    \right],
\end{equation}
where $\gamma$ is the Lorentz factor of the tracer particle and $\mathbf{v}$ is the velocity in the Eulerian grid at the position of the Lagrangian particle and scaled to the speed of light. 

\subsection{Polarisation maps}
\label{sec:polarisation_maps}

\begin{figure}
  \centering
  \includegraphics[width=8cm]{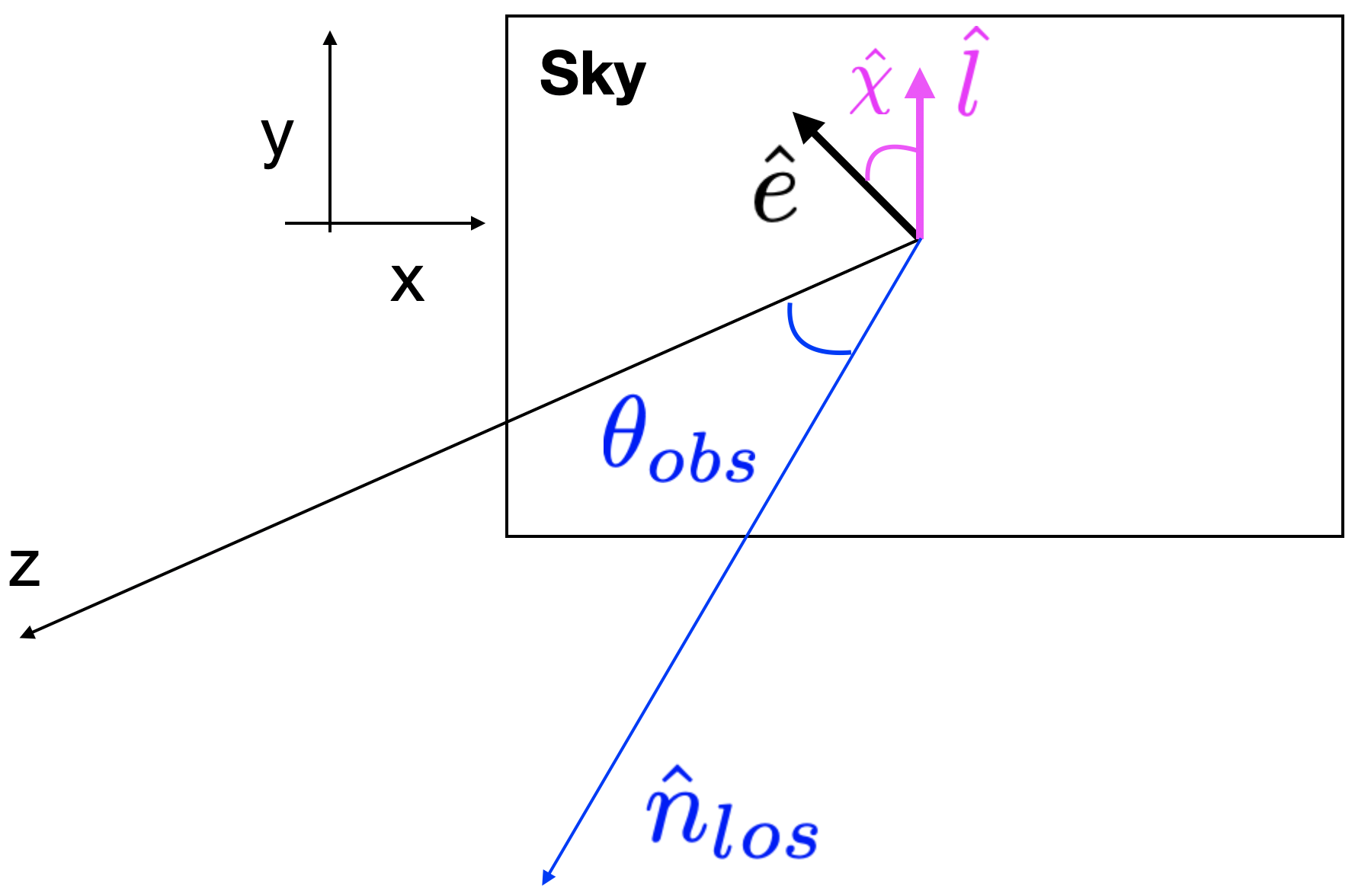}
    \caption{Schematic of the plane of the sky. The two polarisations on the plane of the sky are defined by the electric field unit vector $\mathbf{\hat e}$. The polarisation angle $\hat{\chi}$ is then defined with respect of the unit vector $\mathbf{\hat{l}}$.}
    \label{fig:scheme}
\end{figure}

In this work, we consider an observer's reference frame in which $z$ lies along the LOS $\mathbf{\hat n_{los}}$ and $x$ and $y$ are in the plane of the sky. That is, we choose the vector $\mathbf{\hat n_{los}}$ according to an observing angle $\theta_{\rm obs}$ with respect of the $z$-axis (see Fig.~\ref{fig:scheme}). The Stokes parameters Q and U maps can be obtained in the same fashion as the specific intensity (or surface brightness) maps by integrating along the LOS as
\begin{equation}\label{Q_stokes}
    Q_{\nu} = \int \mathcal{J}_{\rm pol}(\nu_{\rm obs},x,y,z)\cos{2\hat{\chi}}dz,
\end{equation}
\begin{equation}\label{U_stokes}
    U_{\nu} = \int \mathcal{J}_{\rm pol}(\nu_{\rm obs},x,y,z)\sin{2\hat{\chi}}dz,
\end{equation}
where $\hat{\chi}$ is the local \textit{polarisation angle}, that is, the angle of the electric field vector in the plane of the sky measured from a unit vector also defined in the plane of the sky (see Sec.~\ref{sec:polarisation_angle}). This can be computed from the polarised emissivity $\mathcal{J}_{\rm pol}$ that we have as a grid quantity, i.e. we assign the information given by the macro-particles back onto the Eulerian grid. The specific intensity (or surface brightness) maps can be obtained by integrating along a LOS as described in \citetalias{dominguezfernandez2020morphology}:
\begin{equation}\label{eq:intensity}
    I_{\nu} = \int \mathcal{J}_{\rm syn}(\nu_{\rm obs},x,y,z)dz,
\end{equation}
where $\mathcal{J}_{\rm syn}$ is the synchrotron emissivity (see Sec.~3.2 in \citetalias{dominguezfernandez2020morphology}). Finally, Eqs.~(\ref{Q_stokes}) and (\ref{U_stokes}) and (\ref{eq:intensity}) allow us to get the degree of (linear) polarisation, or also referred as the \textit{polarisation fraction}:
\begin{equation}\label{eq:pol_degree}
    \Pi_{\nu} = \frac{\sqrt{Q_{\nu}^2 + U_{\nu}^2}}{I_{\nu}}.
\end{equation}
%

\subsection{The polarisation angle}
\label{sec:polarisation_angle}

The PLUTO code computes the local polarisation angle at each Lagrangian particle following the method described in \citet{2003ApJ...597..998L}. The electric field unit vector of a linearly polarised electromagnetic wave in the comoving frame is normal to the local magnetic field unit vector, $\mathbf{\hat B'}$, and to the LOS, $\mathbf{\hat n_{los}'}$, i.e., directed along the unit vector $\mathbf{\hat e'}= \mathbf{\hat n_{los}'} \times \mathbf{\hat B'}$. In this way, the radiated magnetic field is therefore $\mathbf{\hat b'} = \mathbf{\hat n_{los}'} \times \mathbf{\hat e'}$. The electric field in the observer's frame obtained with Lorentz transformations is
\begin{equation}
    \mathbf{e} = \gamma \left[ \mathbf{\hat e'} -
    \frac{\gamma}{\gamma + 1}(\mathbf{\hat e'} \cdot \mathbf{v}) \mathbf{v} - \mathbf{v} \times \mathbf{\hat b'}
    \right].
\end{equation}
 On the other hand, the Lorentz transformation of Maxwell’s equations in the ideal MHD case $\mathbf{E} + \mathbf{v} \times \mathbf{B} = 0$ allows us to express the comoving frame unit vector $\mathbf{\hat B'}$ in terms of the observer's frame unit vector $\mathbf{\hat B}$:
\begin{equation}\label{eq:B_obsframe}
    \mathbf{\hat B} = \frac{1}{\sqrt{1-(\mathbf{\hat B'} \cdot \mathbf{v})^2}} \left[ \mathbf{\hat B'} -
    \frac{\gamma}{\gamma + 1}(\mathbf{\hat B'} \cdot \mathbf{v}) \mathbf{v}
    \right].
\end{equation}
Using Eqs.~(\ref{eq:nlos_comoving}) and (\ref{eq:B_obsframe}), we obtain a general expression giving the polarisation vector in terms of the observed quantities:
\begin{equation}
    \mathbf{\hat e} = \frac{\mathbf{\hat n_{los}} \times \mathbf{q}}{\sqrt{q^2 -(\mathbf{\hat n_{los}} \cdot \mathbf{q})^2}},
\end{equation}
where

\begin{equation}
    \mathbf{q} = \mathbf{\hat B} + \mathbf{\hat n_{los}} \times (\mathbf{v} \times \mathbf{\hat B}).
\end{equation}
We can then introduce a unit vector $\mathbf{\hat l}$ normal to the plane containing $\mathbf{\hat n_{los}}$ (in our case it is defined in the $y$--axis of plane of the sky), and obtain the components of the $\mathbf{q}$ vector in the plane of the sky:
\begin{equation}\label{eq:cosine}
    \cos \hat{\chi} = \mathbf{\hat e} \cdot (\mathbf{\hat n_{los}} \times \mathbf{\hat l})
\end{equation}
\begin{equation}\label{eq:sine}
    \sin \hat{\chi} = \mathbf{\hat e} \cdot \mathbf{\hat l}.
\end{equation}
Eqs.~(\ref{eq:cosine}) and (\ref{eq:sine}) can be then rewritten in terms of $2\hat{\chi}$ using simple trigonometric relations and finally substituting in Eqs.~(\ref{Q_stokes}) and (\ref{U_stokes}). 
The plane of polarisation suffers additional rotation as the radiation propagates through a magnetised medium, due to Faraday rotation. In order to take this (intrinsic) rotation into account, an additional contribution should be added to the observed  polarisation angle:
\begin{equation}\label{eq:RM_angle}
    \hat{\chi} \longrightarrow \hat{\chi} + \mathrm{RM} \, \lambda_{obs}^2,
\end{equation}
where RM is the Rotation Measure defined as
\begin{equation}
    \mathrm{RM}= 0.812 \, \, \mathrm{rad} \, \mathrm{m}^{-2} \int n_e B_{\parallel} \, dl,
\end{equation}
with the electron density $n_e$ in units of $10^{-3}$cm$^{-3}$, the component of the magnetic field parallel to the LOS $B_{\parallel}$ in $\mu$G and $dl$ in pc. This means that when also considering the intrinsic Faraday rotation of the radio source, we substitute Eq.~(\ref{eq:RM_angle}) into Eqs.~(\ref{Q_stokes}) and (\ref{U_stokes}). We will study this effect in Sec.~\ref{sec:RM}.

Note that while $\hat{\chi}$ defines the intrinsic angle at each point in the 3D simulation box, the polarisation $E$--vector angle as measured by the observer is an integrated quantity which is computed using the Q and U Stokes maps: \footnote{The computation of $\arctan(\theta)$ needs to consider the four quadrants, i.e. $\theta \in [-\pi,\pi]$ }
\begin{equation}\label{eq:pol_angle}
    \psi = \frac{1}{2} \arctan \left( \frac{U_{\nu}}{Q_{\nu}} \right).
\end{equation}
In the following we will use the $\psi$ notation to refer to the polarisation $E$--vector angle.

\begin{figure}
  \centering
  \includegraphics[width=0.7\columnwidth]{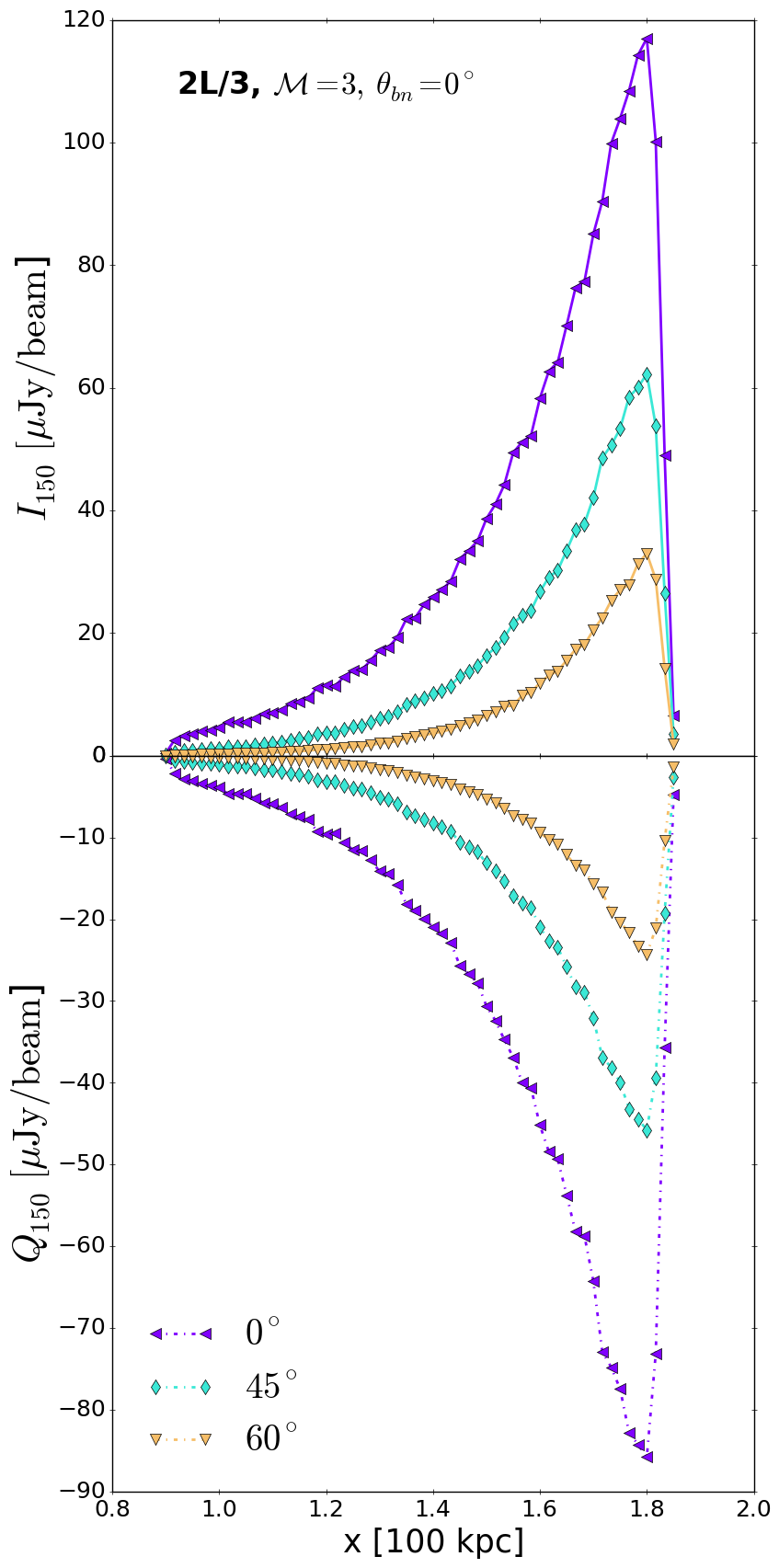}
    \caption{\textbf{Stokes I} parameter (or surface brightness) at 150 MHz (\textit{top panel}) and \textbf{Stokes Q} parameter at 150 MHz (\textit{bottom panel}) integrated along the $z$-axis at 178 Myr. We show the observing angles $\theta_{\rm obs}=0^{\circ}$,$45^{\circ}$ and $60^{\circ}$ in purple, cyan and yellow colours, respectively. We considered a beam of $\theta^2 = 5"\times5"$ to get units of $\mu$Jy/beam in order to have a LOFAR-HBA alike configuration. No smoothing is applied. Note that $\theta_{obs}$ gives the angle between the local magnetic field and the LOS.}
    \label{fig:pol_maps_UNI}
\end{figure}
%

\section{Results}
\label{sec:results2}

\subsection{Uniform medium}
\label{sec:pol_uniform}

%
To test our model, we start by presenting the polarisation of a simple uniform medium. This means that the initial conditions for the pre-shock region (III) are equal to those in pre-shock region (II). We considered two orientations of the magnetic field (see 1 and 2 models in Table \ref{table:init2}): in the $x$--direction parallel to the shock normal, i.e. $\theta_{bn}=0^{\circ}$, and in the $y$--direction perpendicular to the shock normal, i.e. $\theta_{bn}=90^{\circ}$. These runs will be used as control runs for comparison with the turbulent media presented in Sec.~\ref{sec:pol_turb}. We considered three different LOS defined by the observing angle, $\theta_{\rm obs}$: $0^{\circ}$, $45^{\circ}$, $60^{\circ}$ (see Sec.~\ref{sec:pol_methods}).

The first case with a magnetic field in the plane of the sky (only an $x$--component), introduces simplifications such as $| \mathbf{B}' \times \mathbf{\hat n_{los}'}| \propto \cos(\theta_{\rm obs})$ and $\cos \hat{\chi}=0$. This also implies that the Stokes $U_{\nu}$ parameter is zero in Eq.~(\ref{eq:pol_degree}) and that we can get estimates of the resulting $Q_{\nu}$ parameter and polarisation degree $\Pi_{\nu}$ in terms of the values at $\theta_{\rm obs}=0^{\circ}$:
\begin{equation}\label{eq:Q0}
    Q_{\nu}(\theta_{\rm obs}) \sim Q_{\nu}(0^{\circ}) \cos^2(\theta_{\rm obs}),
\end{equation}
and therefore, $\Pi_{\nu}(\theta_{\rm obs})\sim Q_{\nu}(0^{\circ})/I_{\nu}(0^{\circ})$. In Fig.~\ref{fig:pol_maps_UNI}, we show the $I_{150}$ (top panel) and $Q_{150}$ (bottom panel) profiles for a frequency of 150 MHz, at an epoch in which the shock front has reached almost the right end of the simulation box ($\approx 178$ Myr). We show the results for the observing angles $0^{\circ}$, $45^{\circ}$, and $60^{\circ}$, respectively. The Stokes $Q$ parameter should be then smaller than $Q(0^{\circ})$, by a factor of $\sim 1/2$ for the $45^{\circ}$ observing angle and $\sim 1/4$ for the $60^{\circ}$ observing angle (see Eq.~(\ref{eq:Q0})), as shown in Fig.~\ref{fig:pol_maps_UNI}. Note that Eqs.~(\ref{Q_stokes}) and (\ref{U_stokes}) are integrated along the $z$-axis, while the different LOS are defined through the observing angle between the LOS and the $z$-axis.

\begin{figure}
    \centering
    \includegraphics[width=0.7\columnwidth]{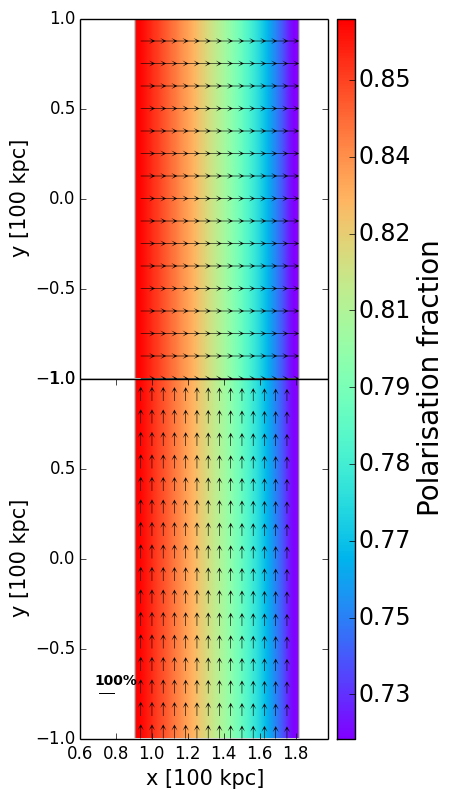}
    \caption{Polarisation fraction for a uniform medium with a magnetic field oriented in the $y$--direction (\textit{upper panel}) and the $x$--direction (\textit{lower panel}). The polarisation fraction computed with Eq.~(\ref{eq:pol_degree}). We overplot the polarisation $E$--vector computed with Eq.~(\ref{eq:pol_angle}).
    }
    \label{fig:pol-uni2}
\end{figure}
\begin{figure}
    \centering
    \includegraphics[width=0.4\textwidth]{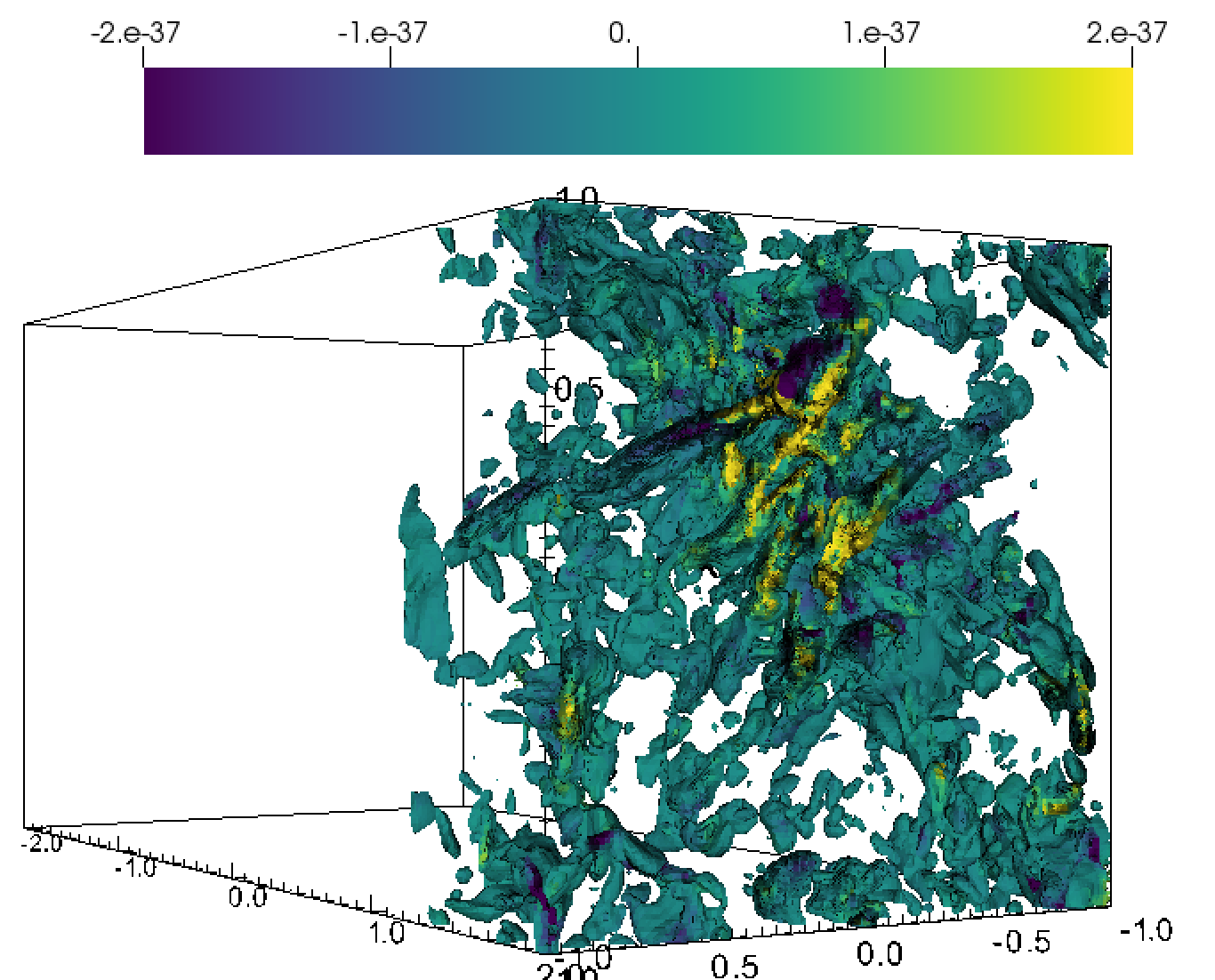}
    \includegraphics[width=0.4\textwidth]{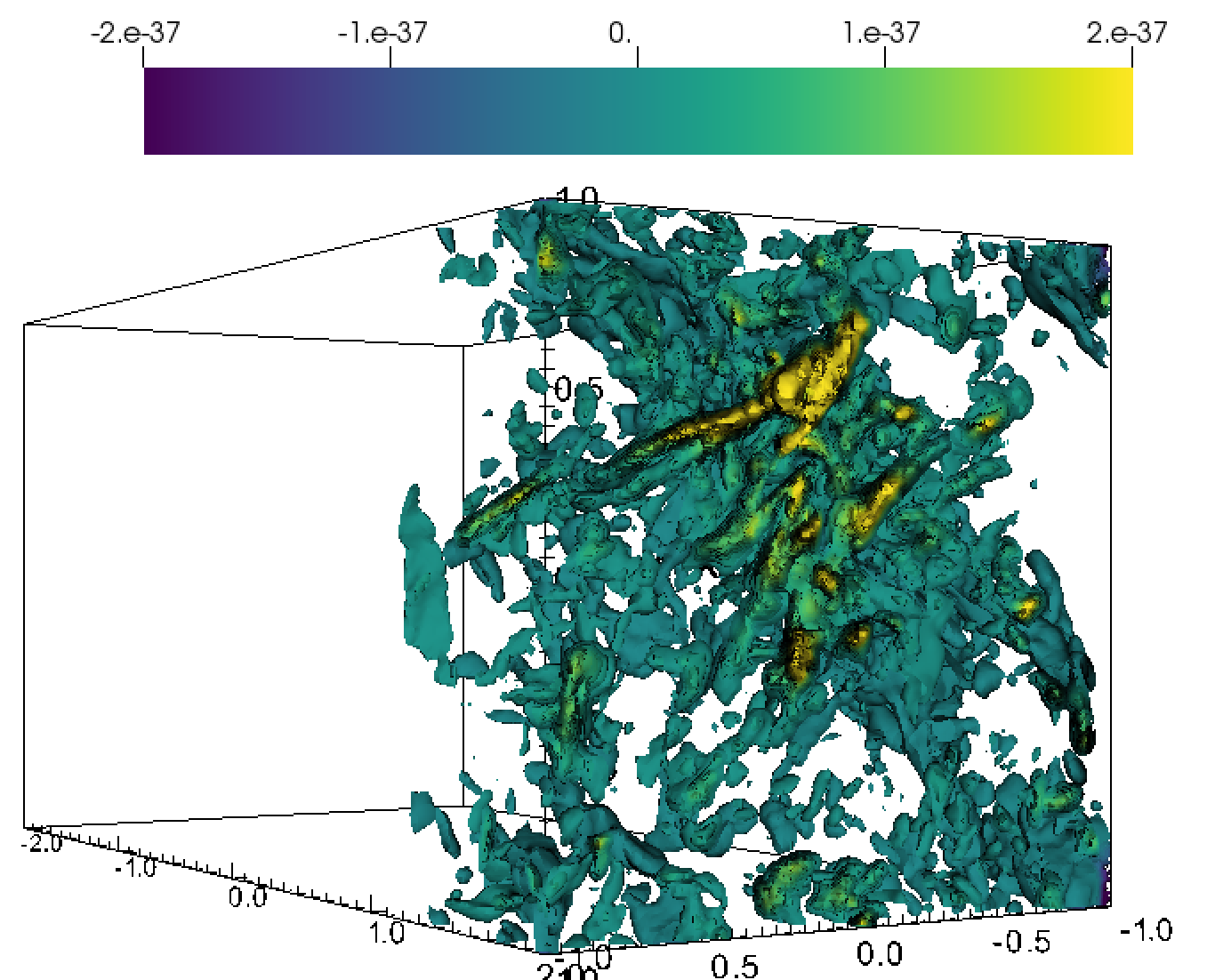}
    \caption{Visualization of the polarised emissivity $\mathcal{J}_{\rm pol}$ isocurves for the $2L/3$, $\mathcal{M}=3$ and $\theta_{bn}=0^{\circ}$ run at $t=178$ Myr and at $\theta_{\rm obs}=60^{\circ}$. The upper panel shows the polarisation corresponding to the \textbf{Stokes Q} parameter (i.e. $\mathcal{J}_{\rm pol}\cos{2\hat{\chi}}$, see Eq.~(\ref{Q_stokes})), while the lower panel shows its corresponding \textbf{Stokes U} parameter (i.e. $\mathcal{J}_{\rm pol}\sin{2\hat{\chi}}$, see Eq.~(\ref{U_stokes})) The emissivity is shown in units of [erg cm$^{-3}$ s$^{-1}$ Hz$^{-1}$ str$^{-1}$]. The axes are shown in units of [100 kpc]. 
    }
    \label{fig:QU60_3d}
\end{figure}
\begin{figure*}
    \centering
    \includegraphics[width=0.9\textwidth]{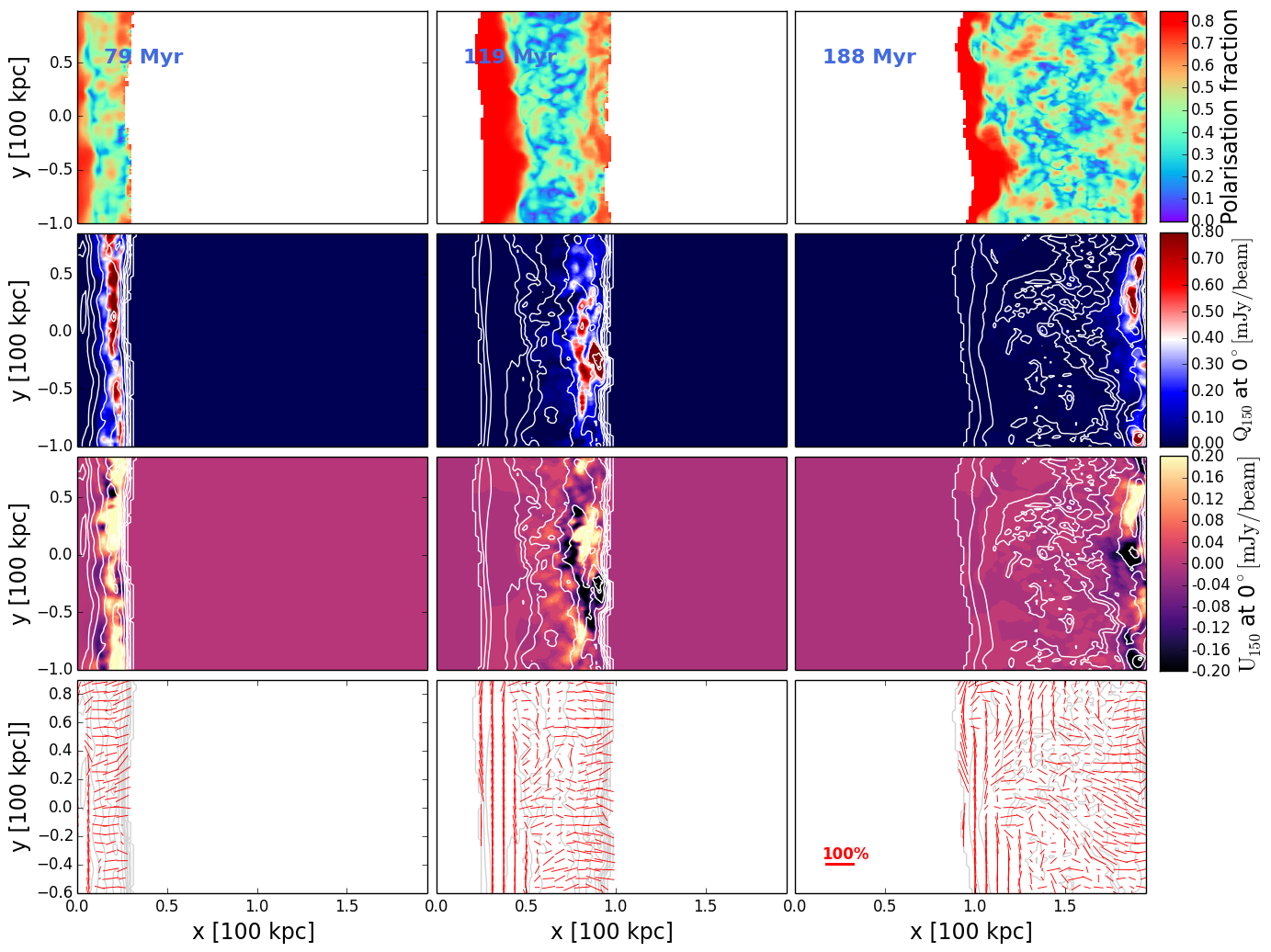}
    \caption{Polarisation of the $2L/3$, $\mathcal{M}=3$ and $\theta_{bn} = 0^{\circ}$ case considering $\theta_{\rm obs}=0^{\circ}$ at 150 MHz. First row: polarisation fraction. Second row: \textbf{Stokes Q} parameter maps. Third row: \textbf{Stokes U} parameter maps. Fourth row: polarisation $E$--vectors computed with Eq.~(\ref{eq:pol_angle}). We overplot the surface brightness contours in all the maps in the second, third and fourth rows. We considered a beam of $\theta^2$ = 6"$\times$6" to obtain the units of mJy/beam.}
    \label{fig:evol}
\end{figure*}
One of the major unknowns in the interpretation of real observed radio relics, is the physical mechanism leading to the high degree of alignment of the magnetic field with the shock front. This can be tested in our simulations.
We compute the polarisation fraction with Eq.~(\ref{eq:pol_degree}) and the polarisation $E$--vector angle with Eq.~(\ref{eq:pol_angle}). 

In Fig.~\ref{fig:pol-uni2}, we show polarisation fraction maps along with the corresponding polarisation $E$-vectors for the two alignments of the magnetic field, i.e. $\theta_{bn}=0^{\circ}$ and $\theta_{bn}=90^{\circ}$. We see a polarisation fraction gradient towards the downstream region of the shock, i.e. the polarisation fraction is lower at the shock front and it increases in the downstream region. The fractional polarisation ranges from $\sim0.7$ to $\sim 0.8$ for the two configurations. This is expected as the freshly activated particles at the shock front age moving downstream, which causes a steepening in their energy spectrum. The polarisation fraction for an injection power-law $E^{-p}$ is: \citep[e.g.][]{1979PhB....30..158E,1979rpa..book.....R}
\begin{equation}\label{eq:pi_power-law}
    \Pi = \frac{p+1}{p+7/3}.
\end{equation}
In our uniform runs, we have $p=2.5$ (corresponding to $\mathcal{M}=3$) corresponding to $\Pi=0.724$ (see shock front values in  Fig.~\ref{fig:pol-uni2}). After activation the energy power-law index, $p$, increases due to synchrotron and inverse Compton losses. One can easily check that if $p$ increases, $\Pi$ increases as well, as shown in Eq.~(\ref{eq:pi_power-law}). In the $\theta_{bn}=0^{\circ}$
case (see lower panel of Fig.~\ref{fig:pol-uni2}), the polarisation $E$-vectors are aligned with the shock surface (or $B$--vector is aligned with the shock normal), while in the $\theta_{bn}=90^{\circ}$
case (see upper panel of Fig.~\ref{fig:pol-uni2}), the polarisation $E$-vectors are aligned with the shock normal as expected. The alignment of the polarisation $E$-vectors with the shock normal is observed, for example, in the relic in MACS\,J1752.0+4440  \citep[e.g.][]{2012MNRAS.426...40B}, the relic in Abell 2744 \citep[e.g.][]{2017ApJ...845...81P}, the Sausage relic \citep{2010Sci...330..347V,2017A&A...600A..18K,2020MNRAS.498.1628L,2021arXiv210206631D}, and the relic in ZwCl\,0008.8+5215 \citep[see][]{2011A&A...528A..38V,2017A&A...600A..18K,2017ApJ...838..110G}.
To summarise, in a uniform medium, we can obtain the same gradient in polarisation for opposite alignments of the magnetic field with respect to the shock normal, i.e. $\theta_{bn}=0^{\circ}$ and $\theta_{bn}=90^{\circ}$.

\subsection{Turbulent medium}
\label{sec:pol_turb}

%
%
An ICM with the level of uniformity presented in Sec.~\ref{sec:pol_uniform} is fairly unrealistic. Therefore, we proceeded to apply the same methods to study more realistic ICM conditions. We present three-dimensional renderings of the polarised emission produced by our modelling for the $2L/3$, $\mathcal{M}=3$ and $\theta_{bn}=0^{\circ}$ run in Fig.~\ref{fig:QU60_3d}, as seen along the LOS defined by $\theta_{\rm obs}=60^{\circ}$ at 150 MHz. In the upper and lower panels of Fig.~\ref{fig:QU60_3d}, we show the polarised emissivity corresponding to the Stokes Q and U parameters, respectively. The polarised emission is not spatially uniform and it coincides with the morphology of the synchrotron emissivity studied in \citetalias{dominguezfernandez2020morphology}. The combination of shock compression and turbulence produces fluctuations in the flow, that are reflected into the shape of threads and filaments. These are visible in the total and polarised intensity maps.  This fining is intriguing as it suggests that such physical configuration (i.e. a shock running over a turbulent ICM) can well explain recently observed radio structures in high-resolution maps of radio relics  \citep[e.g.][]{2018ApJ...852...65R,2020A&A...636A..30R,rajpurohit2020understanding,2014ApJ...794...24O,2017ApJ...835..197V,2018ApJ...865...24D}. The characteristic size of the synchrotron emissivity and the magnetic field were previously studied in \citetalias{dominguezfernandez2020morphology}. There we found that both are of the order of $\sim 70$--100 kpc,  depending on the model (see Fig. 18 in \citetalias{dominguezfernandez2020morphology}). Nevertheless, the size of single threads, bristles and filaments in the synchrotron emissivity are not simply reflecting the shape of magnetic structures. The observed radio emission also depends on the particles' energy spectrum, which differs from location to location due to the different energy losses across the relic area. Likewise, the emissivity fluctuations may be even larger if the acceleration efficiency has a strong dependence on the local plasma conditions. It is possible to measure more precisely the topology of each of these structures using other methods such as the Minkowsky functionals and/or dimensionless
measures such as filamentarity and planarity \citep[e.g.][]{1997ApJ...482L...1S,1999ApJ...526..568S,1998ApJ...495L...5S}. Nevertheless, we leave the topological data analysis for future work.

Next, we show the projected temporal evolution of this same run as viewed from a LOS defined by $\theta_{\rm obs}=0^{\circ}$ in Fig.~\ref{fig:evol}. This figure shows how the polarised emission evolves over a period of $\sim$90 Myr. The three snapshots show how the shock discontinuity is travelling from region II towards the right-hand side of our simulation box. Meanwhile, additional shock-induced turbulence develops in the downstream as the discontinuity crosses region III. The first row shows the polarisation fraction maps as computed with Eq.~(\ref{eq:pol_degree}). The second and third rows show the Stokes Q and U maps used to compute the corresponding polarisation fraction maps, for which we used Eqs.~(\ref{Q_stokes}--\ref{U_stokes}). We also overplot the Stokes I parameter (i.e. surface brightness) contours in white for completeness. Finally, the fourth row shows the polarisation $E$-vectors as computed with the Stokes Q and U maps and Eq.~(\ref{eq:pol_angle}). 

Comparing to the uniform media described in Sec.~\ref{sec:pol_uniform}, we find that in the presence of turbulence, the polarisation fraction does not increase towards the downstream as in the uniform medium, where we observed a gradient.
Instead, a turbulent medium reflects fluctuations in the polarisation fraction maps. The shock front is highly polarised, i.e. up to $\sim 70$\% as the shock sweeps up the transverse components of the magnetic field, effectively aligning the magnetic field vectors. This effect can also be observed in the fourth row of Fig.~\ref{fig:evol}, where the polarisation $E$-vectors at the shock front are mostly aligned with the shock normal. Moreover, downstream of the shock the polarisation fraction fluctuates and can decrease to $\lesssim 10$\%. In particular, the downstream polarisation fraction is affected by magnetic fluctuations in the medium. This is seen in the fourth row of Fig.~\ref{fig:evol}, where the $E$-vectors are more randomly aligned in the downstream region. Note that at the very end of the downstream region, the polarisation fraction is high because this is part of region II with a uniform magnetic field oriented in the $x$--direction (see Sec.~\ref{section:num_set-up2}). The $E$-vectors in this region are perpendicular to the orientation of the magnetic field as shown in Fig.~\ref{fig:pol-uni2}. Therefore, this region will be omitted in the subsequent analysis.

\begin{figure}
    \centering
    \includegraphics[width=0.9\columnwidth]{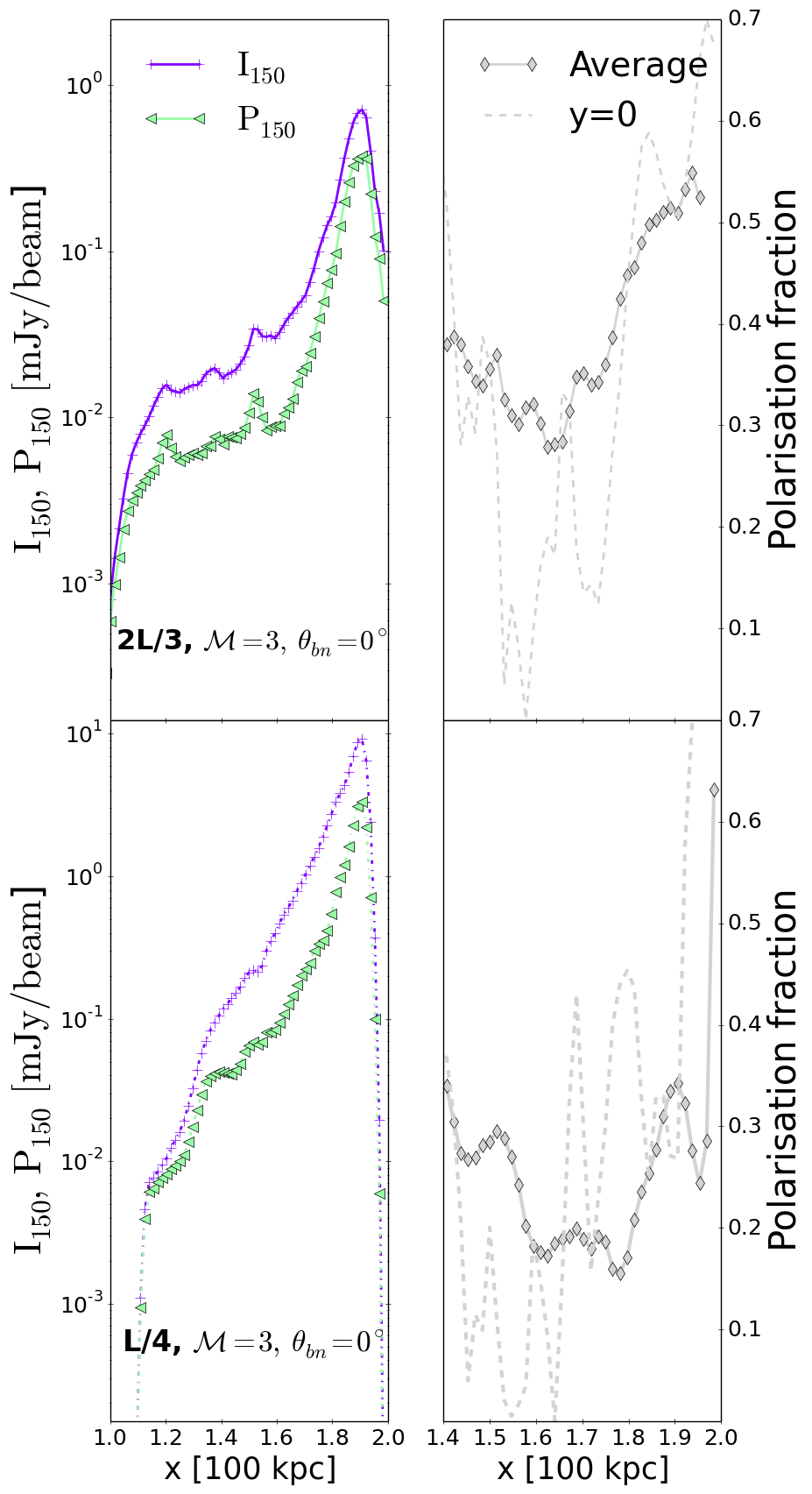}
    \caption{Average 1D surface brightness (\textit{purple crosses}), polarised intensity (\textit{green triangles}) and polarisation fraction (\textit{grey diamonds}) profiles considering $\theta_{\rm obs}=0^{\circ}$ at 150 MHz. We also include a 1D polarisation fraction profile (\textit{dashed grey line}) corresponding to $y=0$ in the polarisation fraction maps in Fig.~\ref{fig:pol_maps_2turb}. The upper panel shows the $2L/3$, $\mathcal{M}=3$ and $\theta_{bn} = 0^{\circ}$ case and the lower panel shows the $L/4$, $\mathcal{M}=3$ and $\theta_{bn} = 0^{\circ}$ case. We considered a beam of $\theta^2$ = 6"$\times$6" to get the units of mJy/beam, but no smoothing is applied.}
    \label{fig:profiles_all}
\end{figure}
\begin{figure}
    \centering
    \includegraphics[width=0.9\columnwidth]{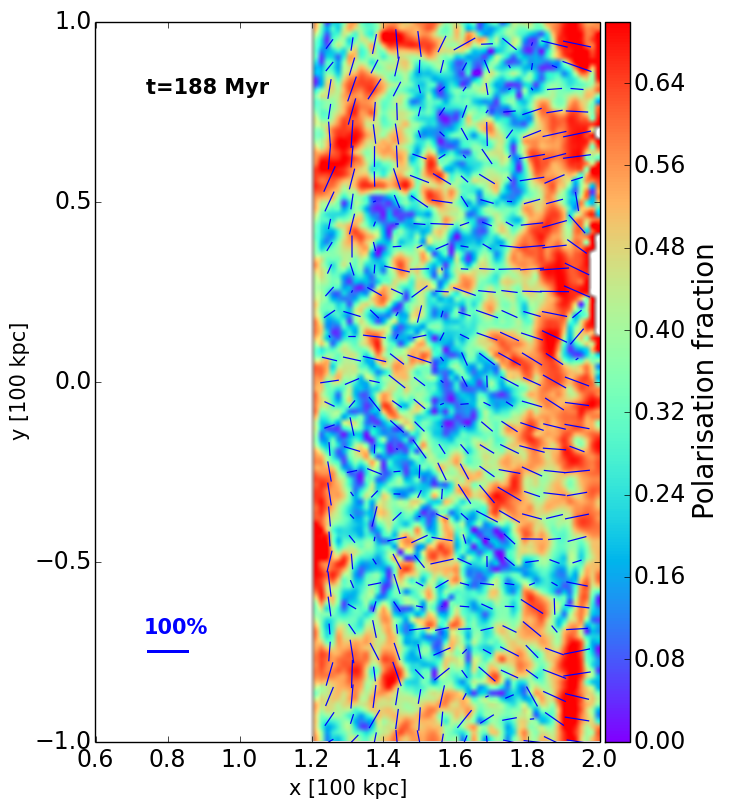}
     \includegraphics[width=0.9\columnwidth]{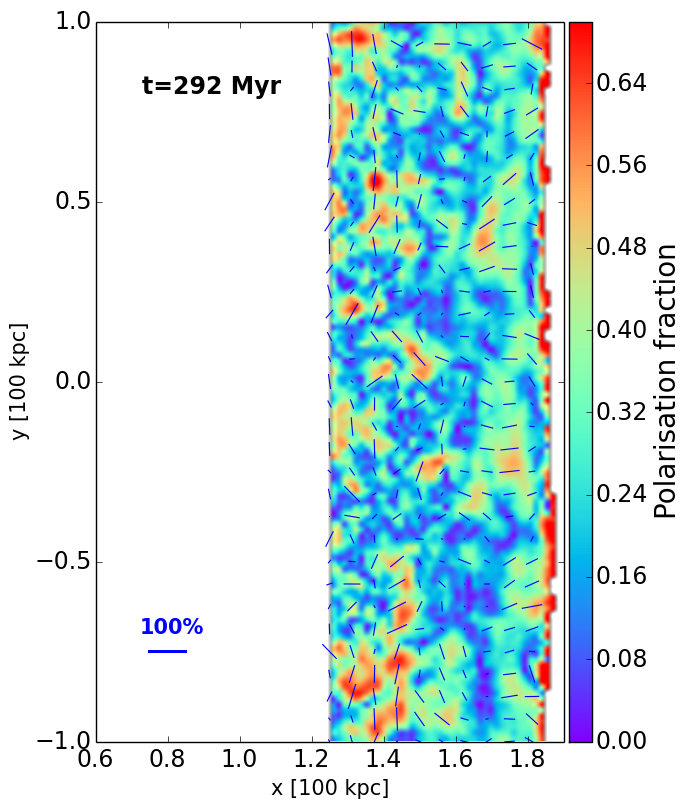}
    \caption{Polarisation fraction maps considering $\theta_{\rm obs}=0^{\circ}$ at 150 MHz. The upper panel shows the $2L/3$, $\mathcal{M}=3$ and $\theta_{bn} = 0^{\circ}$ case and the lower panel shows the $L/4$, $\mathcal{M}=3$ and $\theta_{bn} = 0^{\circ}$ case.}
    \label{fig:pol_maps_2turb}
\end{figure}
\begin{figure}
    \centering
    \includegraphics[width=0.9\columnwidth]{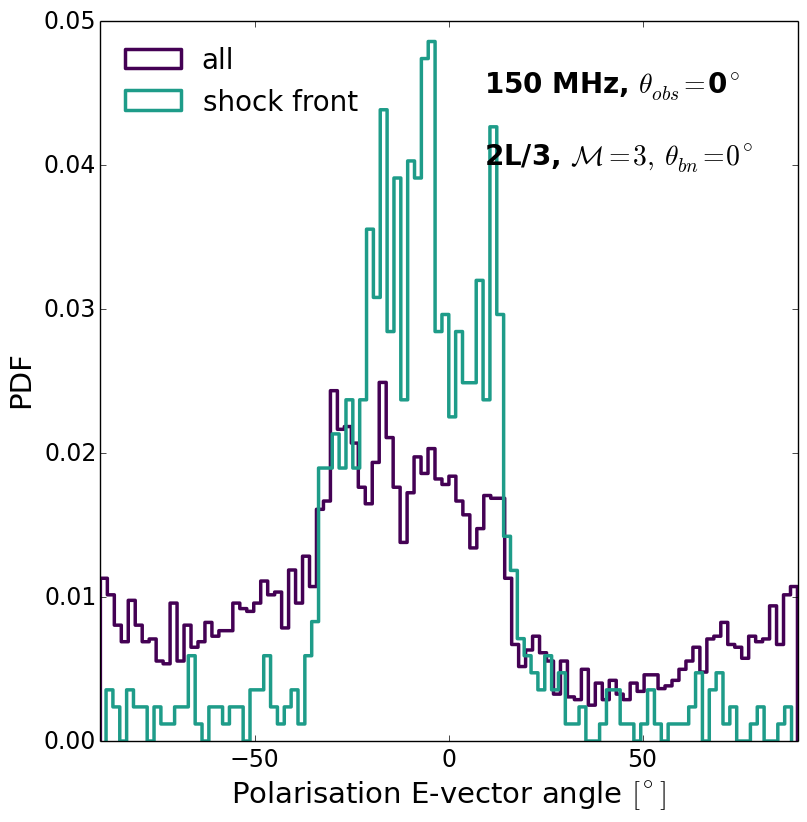}
     \includegraphics[width=0.9\columnwidth]{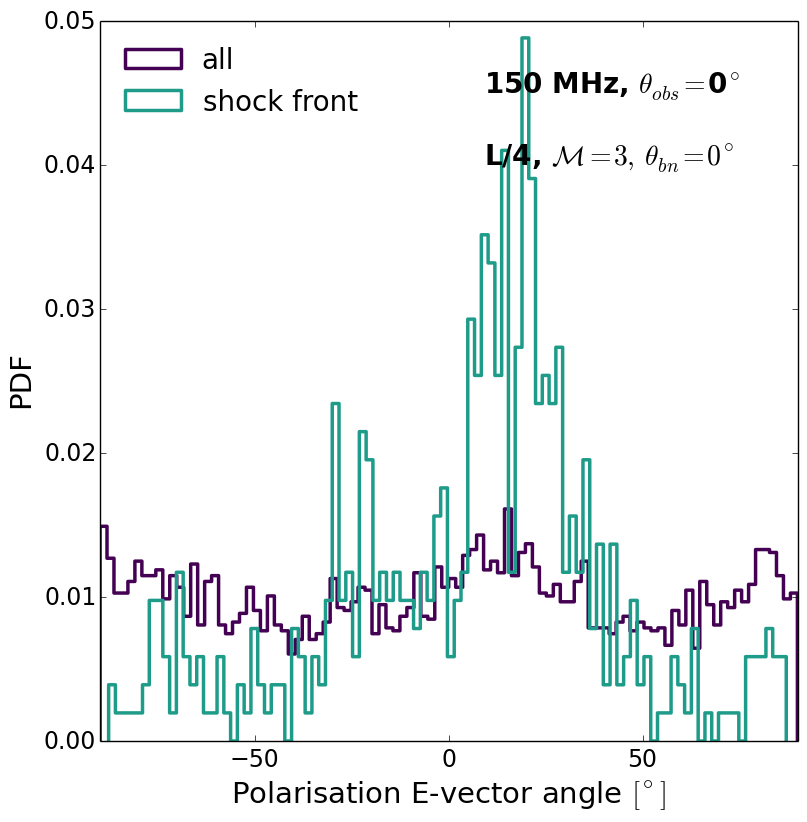}
    \caption{PDF of the polarisation $E$-vector angles as obtained from the maps considering $\theta_{\rm obs}=0^{\circ}$ at 150 MHz. The purple PDF corresponds to a $\sim$ 60 kpc downstream region (see Fig.~\ref{fig:pol_maps_2turb}) and the green PDF is restricted to the shock front region. The upper panel shows the $2L/3$, $\mathcal{M}=3$ and $\theta_{bn} = 0^{\circ}$ case and the lower panel shows the $L/4$, $\mathcal{M}=3$ and $\theta_{bn} = 0^{\circ}$ case.}
    \label{fig:align_150}
\end{figure}
\begin{figure}
    \centering
    \includegraphics[width=0.9\columnwidth]{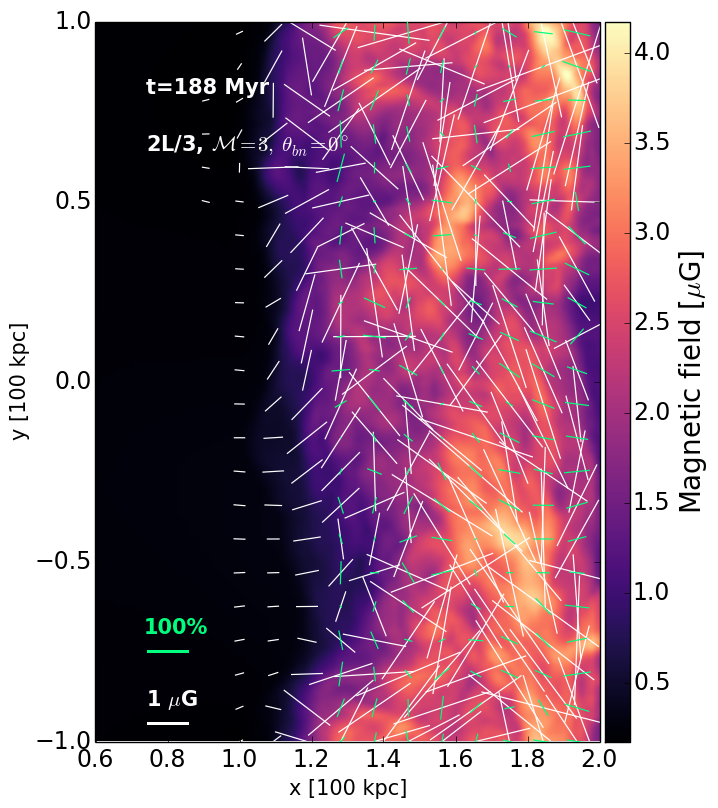}
     \includegraphics[width=0.9\columnwidth]{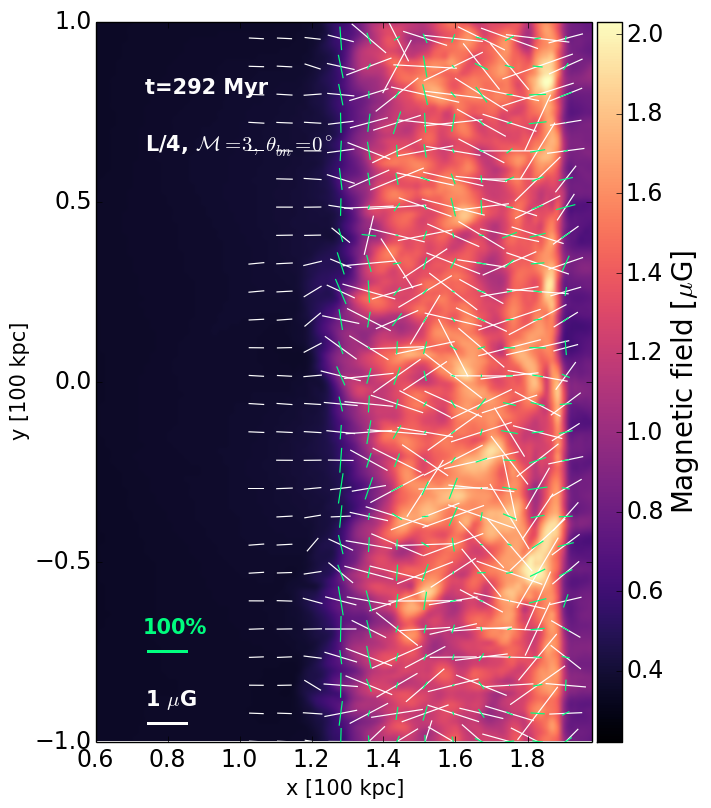}
    \caption{Projected magnetic field of the Eulerian grid. The magnetic field vectors are shown in white and the polarisation $E$-vector in green. The upper panel shows the $2L/3$, $\mathcal{M}=3$ and $\theta_{bn} = 0^{\circ}$ case and the lower panel shows the $L/4$, $\mathcal{M}=3$ and $\theta_{bn} = 0^{\circ}$ case.}
    \label{fig:maps_B}
\end{figure}

In Fig.~\ref{fig:profiles_all} we show 1D profiles of the surface brightness, polarised intensity and polarisation fraction for the two different turbulent media: $2L/3$, $\mathcal{M}=3$ and $\theta_{bn}=0^{\circ}$ and $L/4$, $\mathcal{M}=3$ and $\theta_{bn}=0^{\circ}$. We show the average over the $y$-axis for the three quantities and we also show the 1D profile of the polarisation fraction selected at a fixed $y$-position (we selected $y=0$). The polarised intensity is computed using the following equation:
\begin{equation}\label{eq:pol_intensity}
    P_{\nu} =\sqrt{Q_{\nu}^2 + U_{\nu}^2}.
\end{equation}
In both cases, the surface brightness profile is smoother and has higher values than the polarised intensity. This highlights the role of the polarisation vectors in the plane of the sky as computed with Eqs.~(\ref{Q_stokes}) and (\ref{U_stokes}). Due to presence of post-shock turbulence, the profile of the  polarisation fraction varies considerably in the downstream region. For a fixed $y$-position, we note that while the shock front reaches a polarisation fraction of $\sim 70$\%, the far downstream region varies from $\lesssim 10$\% to 57\% in the $2L/3$, $\mathcal{M}=3$ and $\theta_{bn}=0^{\circ}$ case and from $\lesssim 10$\% to $\lesssim 50$\% in the $L/4$, $\mathcal{M}=3$ and $\theta_{bn}=0^{\circ}$ case. A high polarisation fraction of order $\sim$ 60\% at the shock front is found in both models. In the downstream regions, this fraction decreases. This last result is in agreement with what has been recently observed in the Sausage relic \citep{2021arXiv210206631D} where the outermost edge reaches intrinsic polarisation values up to $\sim 60$\% and then the fraction decreases towards the downstream.

The discrepancies between both turbulent media are almost indistinguishable in the polarisation profiles, but there are differences that can be better observed in the polarisation fraction maps shown in Fig.~\ref{fig:pol_maps_2turb}. Again we have excluded the uniform medium in region II. While both turbulent media can have a high polarisation fraction at the shock front, the depolarisation in the downstream region (considering $\sim 60$ kpc distance from the shock front) is higher for the $L/4$ case than for the $2L/3$ case. This is expected as smaller magnetic fluctuations will cause more depolarisation (see also Fig.~\ref{fig:profiles_all}). This result is linked to the evolution of the magnetic power spectrum. In particular, the characteristic length of the magnetic power spectrum remains smaller in the $L/4$ case compared to that of the $2L/3$ case during the whole evolution (see $\lambda_B$ in Fig. 18 in \citetalias{dominguezfernandez2020morphology}). This means that the $2L/3$ magnetic field stays more coherent at larger scales than the $L/4$ case.
Similar polarisation profiles have been observed in radio relics, most clearly in the Toothbrush relic \citep[e.g.][]{2012A&A...546A.124V,2020A&A...642L..13R} and in MACSJ0717.5+3745 relic (e.g. \citealt{2009A&A...503..707B}, {\color{blue} Rajpurohit et al. submitted}). Another example where fractional polarisation goes up to 70\% is the large relic in the galaxy cluster Abell 2256 \citep[e.g.][]{2014ApJ...794...24O}. However, the relic in Abell 2256 is more complex and cannot be directly compared to our simulations.
\begin{figure}
    \centering
    \includegraphics[width=0.9\columnwidth]{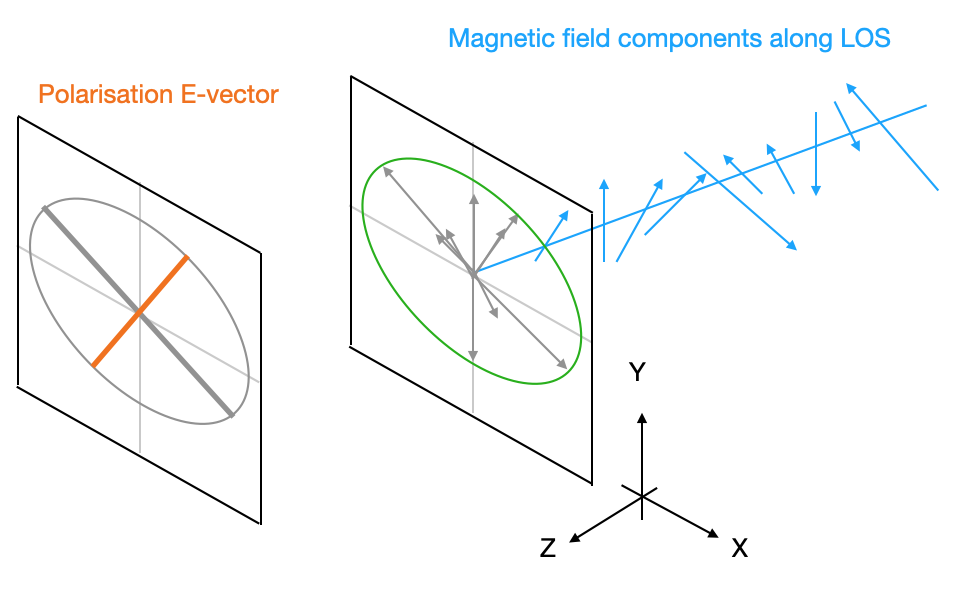}
    \caption{Schematic of different magnetic field components that contribute to one LOS as seen from the $z$-axis. If the final distribution of the magnetic field components is anisotropic with a preferred direction, then we would observe an ellipse in the $B_x$-$B_y$ plane. The alignment of the polarisation $B$-vector ($E$-vector) is defined then by the direction of the major (minor) radius of the ellipse.}
    \label{fig:sketch_align}
\end{figure}
\begin{figure}
    \centering
    \includegraphics[width=0.9\columnwidth]{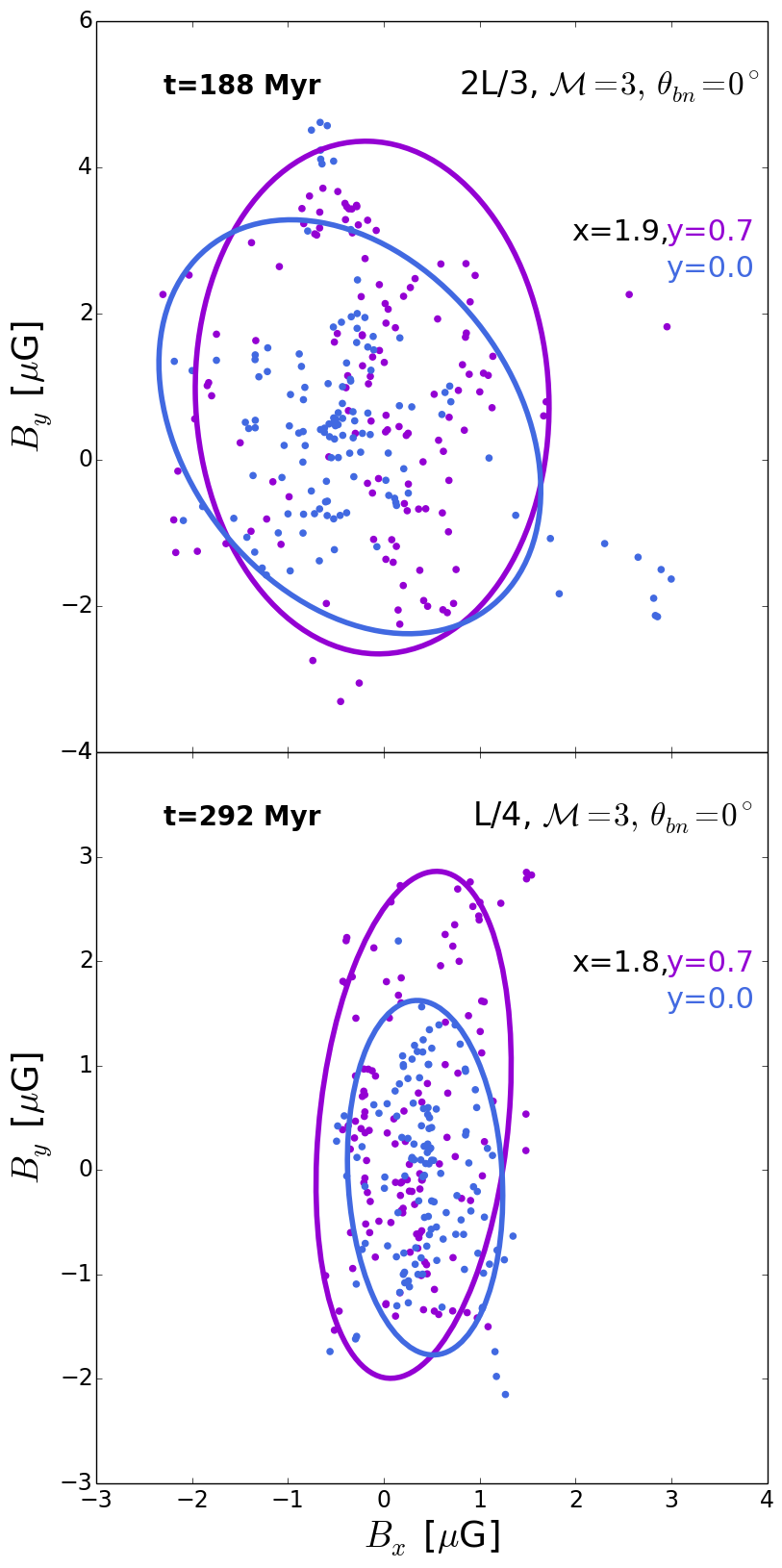}
    \caption{$By$-$Bx$ distribution of magnetic field components along two LOS. The LOS are taken at a fixed box coordinates $x$ and $y$ (see e.g. Fig.~\ref{fig:pol_maps_2turb}).  \textit{Upper panel}: $2L/3$, $\mathcal{M}=3$ and $\theta_{bn} = 0^{\circ}$ case. \textit{Lower panel}: $L/4$, $\mathcal{M}=3$ and $\theta_{bn} = 0^{\circ}$ case. The $x$ coordinate is located at the position of the shock front and we show two different LOS by selecting two $y$ positions (purple and blue). A $3\sigma$ covariance confidence ellipse is shown in each plot.}
    \label{fig:components_LOS}
\end{figure}
\begin{figure*}
    \centering
    \includegraphics[width=0.29\textwidth]{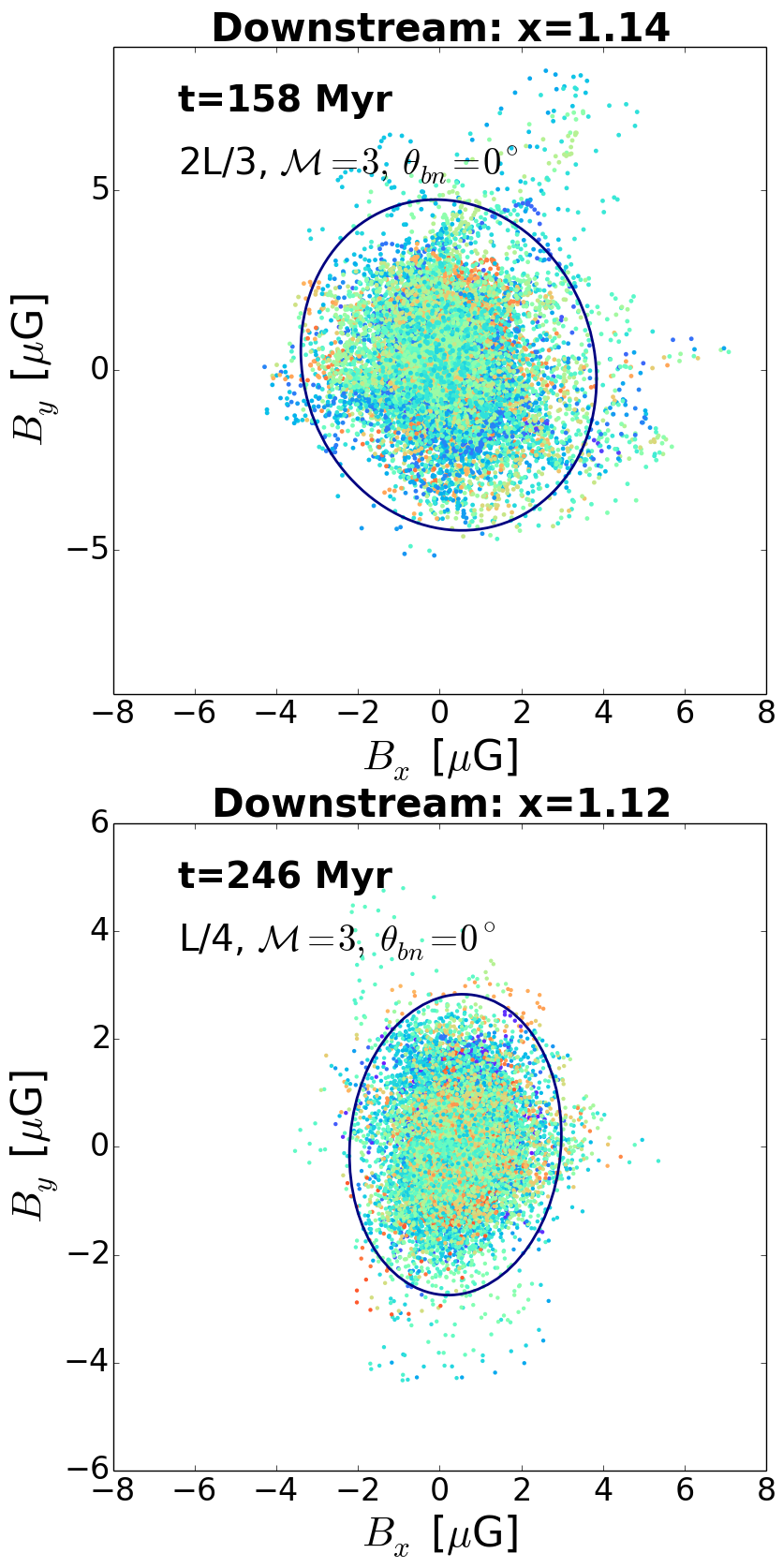}
    \includegraphics[width=0.345\textwidth]{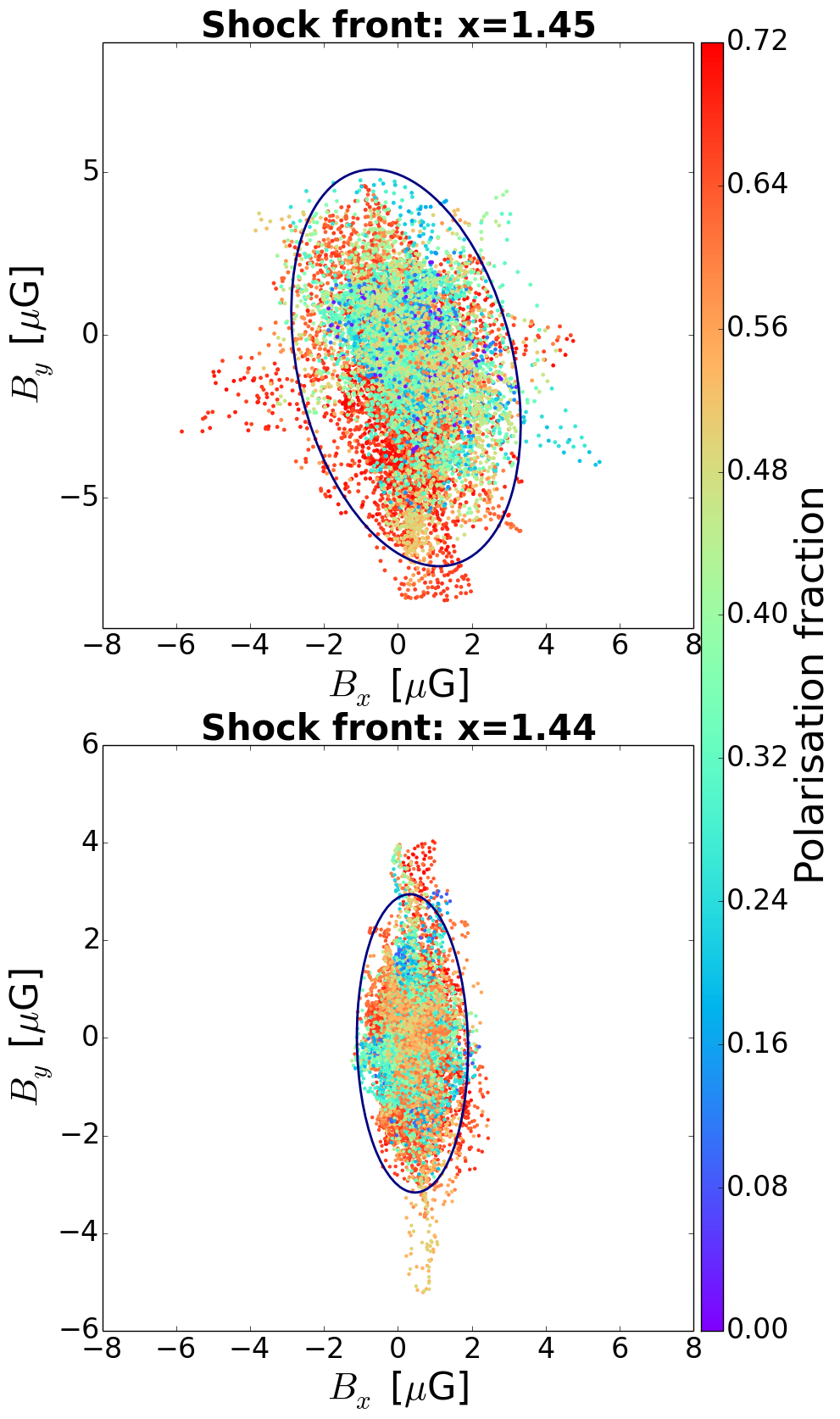}
    \includegraphics[width=0.29\textwidth]{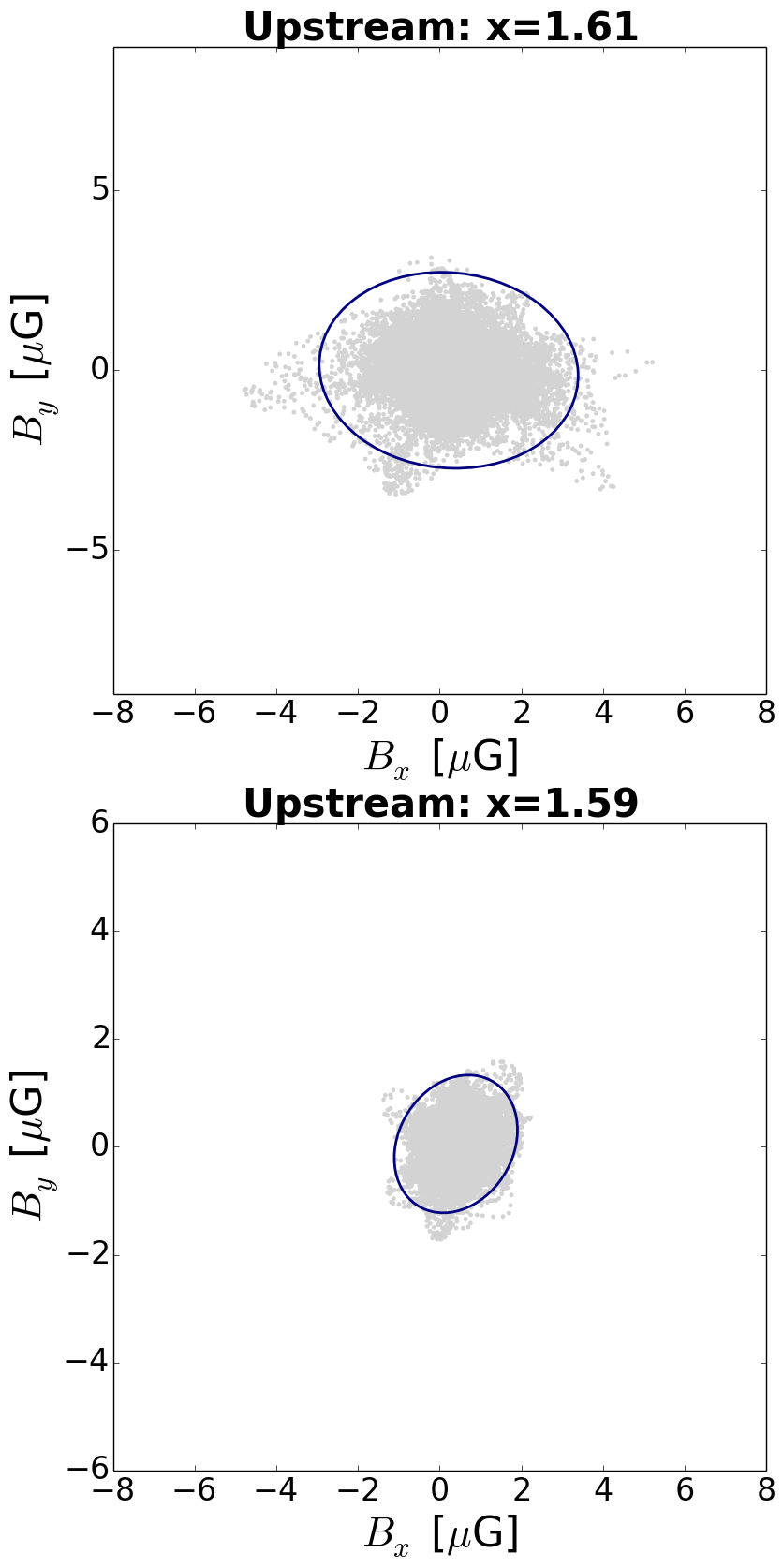}
    \caption{$Bx$-$By$ distribution of magnetic field components along various LOS (similar to Fig.~\ref{fig:components_LOS}). In this case, we fix only the $x$ position and we present all the different LOS along the $y$-axis. The first, second and third columns correspond to a specific $x$ position at the downstream, shock front and upstream, correspondingly. \textit{Upper row}: $2L/3$, $\mathcal{M}=3$ and $\theta_{bn} = 0^{\circ}$ case. \textit{Lower row}: $L/4$, $\mathcal{M}=3$ and $\theta_{bn} = 0^{\circ}$ case. The scatter plots presented in the first two columns are coloured according to the value of the polarisation fraction in the polarisation fraction maps at every ($x,y$) position (see Fig.~\ref{fig:pol_maps_2turb}). Since the upstream is not polarised, we present the distribution without colouring. A $3\sigma$ covariance confidence ellipse is shown in each of the distributions.}
    \label{fig:B-field_downstream}
\end{figure*}
\begin{figure}
    \centering
    \includegraphics[width=0.9\columnwidth]{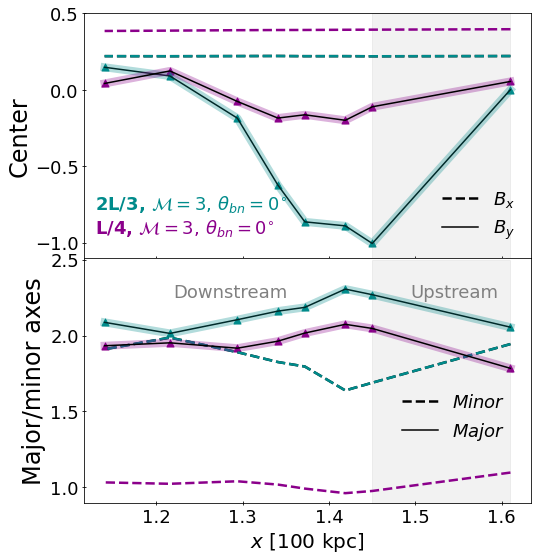}
    \caption{Centre and minor and major axes of the covariance confidence ellipse fitted to the $B_x$-$B_y$ distributions as seen in Fig.~\ref{fig:B-field_downstream} in the upstream, shock front and downstream regions. \textit{Upper panel}: Mean given by the ellipse centre. The dashed (solid) lines correspond to the mean values of the $B_x$ ($B_y$) component. \textit{Lower panel}: Minor (dashed lines) and major (solid lines) axes of the fitted ellipse. The values are given in units of $\mu$G.}
    \label{fig:ellipse_B}
\end{figure}

In Fig.~\ref{fig:pol_maps_2turb} we also overplotted the polarisation $E$-vectors. Both models produce polarisation $E$-vectors aligned with the shock normal at the position of the shock front.
In order to visualise the overall alignment of the polarisation $E$-vectors, we produce PDFs of the polarisation $E$-vector angle $\psi$. In Fig.~\ref{fig:align_150}, we show the PDF of these polarisation $E$-vectors for both turbulent media (see corresponding Fig.~\ref{fig:pol_maps_2turb}). We also compute the PDF corresponding only to the cells tracking the shock front (see \citetalias{dominguezfernandez2020morphology}). This is possible because our numerical implementation relies on tracking the shock discontinuity where the CRe particles are being activated at each time-step. In Fig.~\ref{fig:align_150} we show the normalised PDF of the $\sim$60 kpc downstream region in purple, and that of the shock front region in green. The PDFs show a similar pattern for both turbulent media. The PDF in the shock front peaks at lower polarisation angles, i.e. $\mid \psi \mid \leq 30^{\circ}$. The $L/4$, $\mathcal{M}=3$ and $\theta_{bn}=0^{\circ}$ case shows a distribution that is narrower than the $2L/3$, $\mathcal{M}=3$ and $\theta_{bn}=0^{\circ}$ case. Remembering that $\psi\sim 0^{\circ}$ is the direction of a polarisation $E$-vector parallel to the initial shock normal, Fig.~\ref{fig:align_150} shows that in a turbulent ICM the polarisation $E$-vectors tend to align with the shock normal.

Comparing our two turbulent models, we find that the $2L/3$ turbulence model produces a broader PDF peaking at $\psi\sim 0^{\circ}$ than the $L/4$ turbulence model. This implies that the final degree of alignment of  polarisation $E$--vectors at the shock front depends on the distribution of upstream turbulent modes. 
In particular, these differences stem from the final magnetic power being higher in the $2L/3$ case than in the $L/4$ case (see magnetic power spectra in Figs. 16 and 17 in \citetalias{dominguezfernandez2020morphology}), and from the larger characteristic scale of the magnetic field in the $2L/3$ case compared to the $L/4$ case (see $\lambda_B$ in Fig. 18 in \citetalias{dominguezfernandez2020morphology}). In \citetalias{dominguezfernandez2020morphology}, we also found that the standard deviation of the perpendicular component of the magnetic field (with respect to the shock normal), i.e. $B_{\perp}=\sqrt{B_y^2 + B_z^2}$, is modified more than the parallel component due to the shock crossing. Specifically, a shock with strength $\mathcal{M}=3$ induces a higher increase in the standard deviation of the perpendicular magnetic component of the $2L/3$ case (31\% increase) than in the L/4 case (7\%). 

\begin{figure}
    \centering
    \includegraphics[width=0.9\columnwidth]{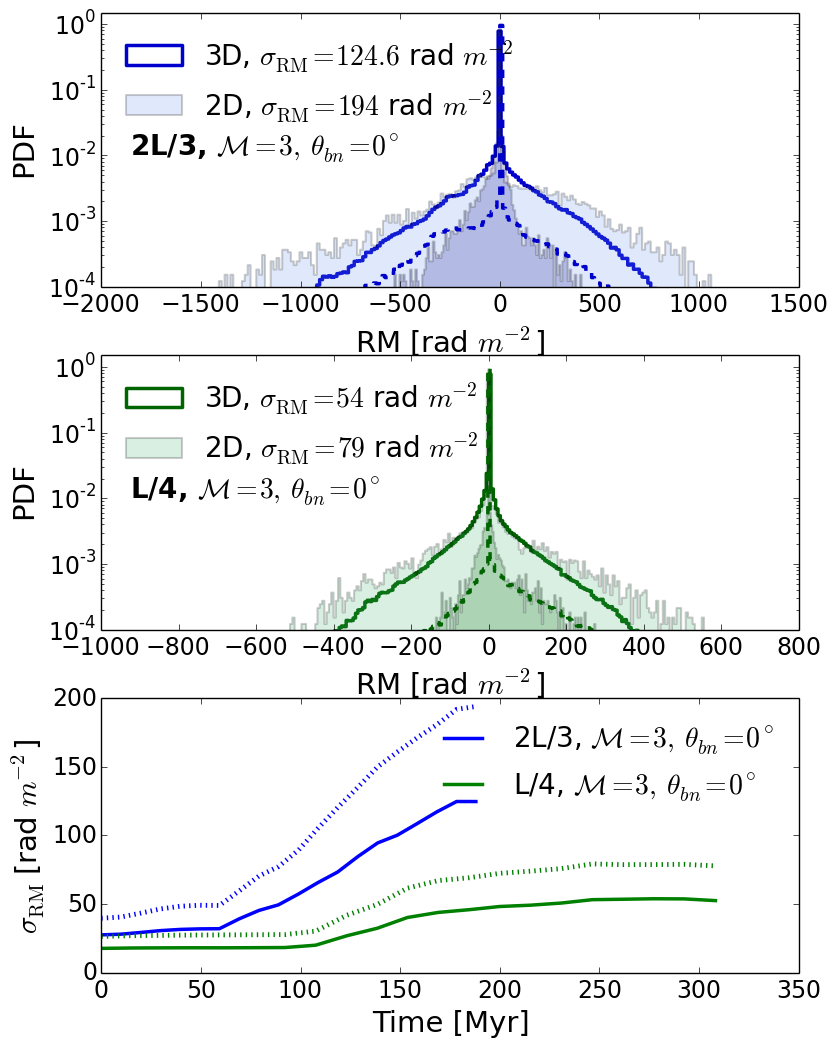}
    \caption{ PDF of the RM for both turbulent media integrated along the $z$-axis: at $t=188$ Myr for the $2L/3$ case (\textit{upper panel}) and at $t=292$ Myr for the $L/4$ case (\textit{middle panel}). The step PDF corresponds to the 3D distribution and the step-filled PDF corresponds to the 2D distribution as integrated over the whole box (200 kpc) along $z$. The dashed lines correspond to the 3D weighted distribution and the PDFs with higher less transparency correspond to the 2D weighted distribution (see Eqs.~\ref{eq:RM_weight1}--\ref{eq:RM_weight2}). \textit{Lower panel}: Evolution of the standard deviation $\sigma_{\mathrm{RM}}$ of the 3D (solid lines) and 2D (dotted lines) distribution for both media. 
    }
    \label{fig:PDF_RM}
\end{figure}

\begin{figure}
    \centering
    \includegraphics[width=0.9\columnwidth]{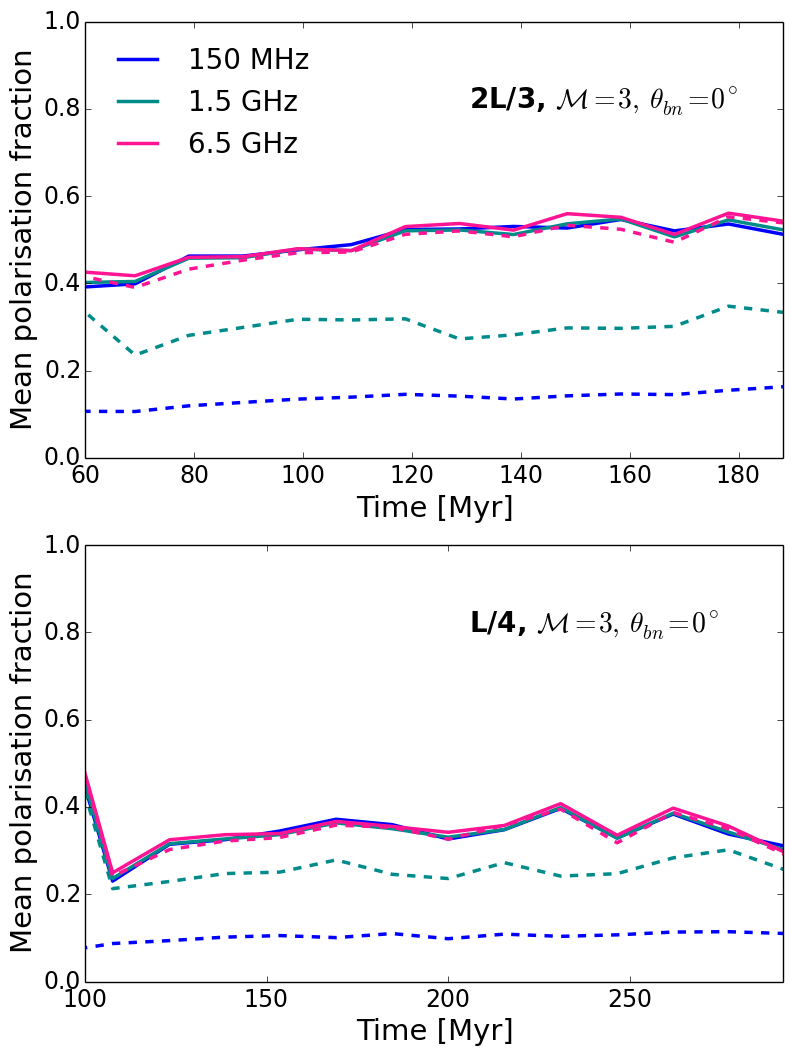}
    \caption{ Evolution of mean polarisation fraction at the shock front for the $2L/3$, $\mathcal{M}=3$ and $\theta_{bn} = 0^{\circ}$ case (\textit{upper panel}) and the $L/4$, $\mathcal{M}=3$ and $\theta_{bn} = 0^{\circ}$ case (\textit{lower panel}). The \textit{solid lines} correspond to the polarisation fraction without RM contribution and the \textit{dashed lines} correspond to the polarisation fraction with RM contribution. Note that we show the time evolution since the shock front enters the turbulent region III.
    }
    \label{fig:Polfrac_RM}
\end{figure}

The orientation or alignment of the polarisation $E$-vectors usually gives us an idea on the degree of alignment of the magnetic field. It is therefore also useful to visualise the magnetic field of the Eulerian grid. We show the projected magnetic field weighted by the synchrotron emissivity,
\begin{equation}\label{eq:B_proj}
    B_{\rm proj} = \frac{\int B \, \mathcal{J}_{\mathrm{syn}} dz}{\int dz},
\end{equation}
along with the corresponding magnetic field vectors in Fig.~\ref{fig:maps_B}. The angle of the magnetic field vector was computed by first projecting the $x$- and $y$--components as in Eq.~(\ref{eq:B_proj}) and finally computing\footnote{Here also $\arctan(\theta)$ considers the four quadrants, i.e. $\theta \in [-\pi,\pi]$ }
\begin{equation}
    \psi_B = \arctan \left( \frac{B_y}{B_x} \right) .
\end{equation}
 We remind the reader that the projected magnetic field vectors as shown in Fig.~\ref{fig:maps_B} may not be identical to the polarisation $B$-vectors. In order to understand the orientation of the polarisation $B$- and $E$-vectors, we show a sketch in Fig.~\ref{fig:sketch_align} explaining how the individual magnetic field components along one particular LOS contribute to the final orientation. There are two possibilities leading to polarisation:
\begin{itemize}
    \item[i)]  \textit{A uniform homogeneous magnetic field}: here all the magnetic field components along the LOS would contribute to the same direction. 
    \item[ii)] \textit{The compression of a tangled magnetic field}: here an anisotropic distribution of the magnetic field components (in the $B_x$-$B_y$ plane in our case) leads to a specific alignment of the polarisation vectors \citep[see also][]{1966MNRAS.133...67B}. Without compression, a randomly oriented magnetic field that produces a perfectly isotropic distribution in the $B_x$-$B_y$ plane will not produce polarisation.
\end{itemize}

The first possibility was already shown in Sec.~\ref{sec:pol_uniform}, while the second possibility explains the results shown in this section. In Fig.~\ref{fig:components_LOS} we show the distributions in the $B_x$-$B_y$ plane for our two turbulent models by considering two particular LOS. Each LOS was selected to be at the location of the shock front (see box coordinates in Fig.~\ref{fig:components_LOS}). The distributions are in general not isotropic due to the compression of the shock. The $B_x$-$B_y$ distribution can be seen as bivariate data with a covariance matrix describing an ellipse.  The eccentricity of the covariance ellipse is related to the anisotropy of the distribution. We therefore characterise each $B_x$-$B_y$ distribution by computing a $3\sigma$ covariance confidence ellipse (see contours in Fig.~\ref{fig:components_LOS} and Fig.~\ref{fig:B-field_downstream}).
The more anisotropic the distribution is, the more elongated the covariance ellipse will be. The alignment of the ellipse major axis then defines the alignment of the polarisation $B$-vector and as a result, the minor axis defines the alignment of the polarisation $E$-vector. Due to beam smearing, we would always have contamination coming from various LOS in observations. It is therefore useful to visualise the $B_x$-$B_y$ distribution for a region instead of a single LOS. In Fig.~\ref{fig:B-field_downstream} we show the corresponding distributions for a fixed $x$-coordinate and consider all the contributions from all the $y$-coordinates. In this case, we show the $B_x$-$B_y$ distribution for regions in the upstream ($\sim$ 30 kpc ahead of the shock front), at the shock front and in the downstream ($\sim$ 20 kpc behind of the shock front) for our two turbulent models at a different simulation time\footnote{Note that the simulation times selected for Fig.~\ref{fig:components_LOS} correspond to the same simulation times discussed previously in this section where the shock front has reached the right computational boundary, whereas the simulation times selected for Fig.~\ref{fig:B-field_downstream} correspond to an earlier phase of the shock propagation which allow us to study the upstream $B_x$-$B_y$ distribution.} (see text in each panel of Fig.~\ref{fig:B-field_downstream}). 
Both turbulent models show similar trends, the $B_x$-$B_y$ distribution is more anisotropic at the shock front and then it becomes more isotropic further in the downstream, i.e. the emission is less polarised in the downstream. The anisotropy at the shock front is observed as a more elongated ellipse in the $B_y$ component due to the fact that the shock compression affects only the components that are perpendicular to the shock normal (i.e. $B_y$ and $B_z$ in our set-up). This explains the overall alignment of the polarisation $E$-vectors in the direction of the shock normal (i.e. $x$-direction in our set-up) at the shock front. Each LOS in Fig.~\ref{fig:B-field_downstream} is coloured according to the polarisation fraction value at every ($x,y$) position in the corresponding polarisation fraction map. One can observe that the highest polarisation fraction is indeed observed at the shock front. This confirms our obtained trends in the polarisation fraction maps as well as the alignment of the polarisation $E$-vectors (see Figs.~\ref{fig:pol_maps_2turb} and \ref{fig:maps_B}). 

Having characterised the $B_x$-$B_y$ distribution with the covariance ellipse,
we can further investigate the expected decrease of anisotropy towards the downstream for each turbulent model compared
to the upstream distribution. In Fig.~\ref{fig:ellipse_B} we show the characteristics of the covariance ellipses as shown in Fig.~\ref{fig:B-field_downstream}. The ellipse centre corresponds to the mean of the distribution, whereas the extent of the ellipse axes defines the anisotropy of the distribution. The way the distribution becomes more isotropic towards the downstream can be mainly seen in the lower panel of Fig.~\ref{fig:ellipse_B} where the size of both axes start to have more similar sizes towards the downstream. In other words, the major axis of the ellipse will start to decrease towards the downstream up to the point where its size is comparable with the minor axis. The $2L/3$ model shows more magnetic amplification than the $L/4$ model as can be observed in the upper panel of Fig.~\ref{fig:B-field_downstream} and in agreement with Fig. 8 of \citetalias{dominguezfernandez2020morphology}. On the other hand, due to this higher magnetic amplification, the $B_x$-$B_y$ distribution will become isotropic at a shorter distance from the shock front in the $2L/3$ case than in the $L/4$ case 
(see lower panel in Fig.~\ref{fig:ellipse_B}). This clearly shows that the degree of depolarisation downstream of the shock depends on the properties of the underlying turbulent medium. Hence, it can inform us also on the properties of the ICM upstream of the shock.

In summary, the shock compression of a turbulent ICM leads to specific characteristics in polarisation: 1) a high polarisation fraction at the shock front and a decrease towards the downstream region; and 2) an alignment of the polarisation $E$--vector with the shock normal. The depolarisation in the downstream region is subject to the properties of the underlying pre-shock turbulent medium. In particular, it is subject to how the $B_x$-$B_y$ distribution becomes more isotropic towards the downstream region. We find that the 2L/3 case, tends to isotropise the $B_x$-$B_y$ distribution at a distance closer to the shock front than the L/4 case.

\begin{figure}
    \centering
    \includegraphics[width=0.8\columnwidth]{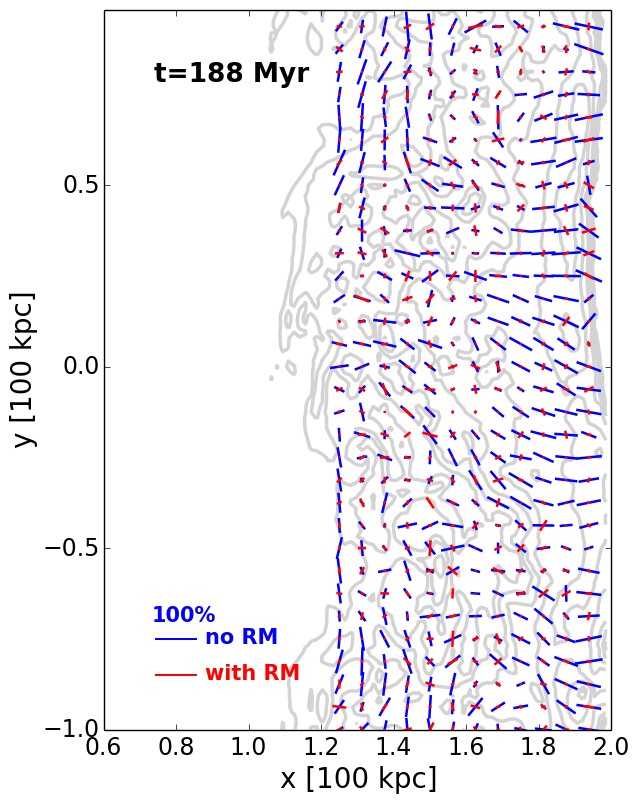}
     \includegraphics[width=0.8\columnwidth]{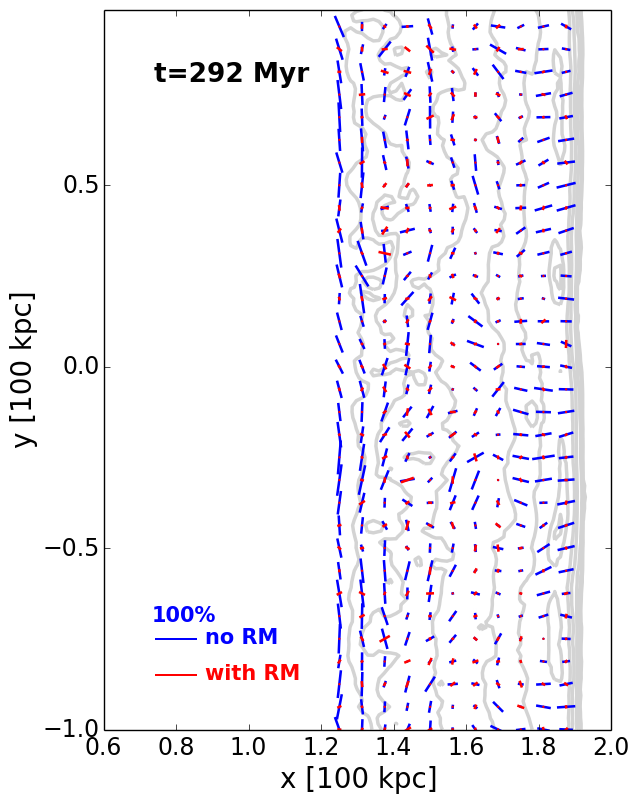}
    \caption{Polarisation E-vectors with (red) and without (blue) the RM contribution considering $\theta_{\rm obs}=0^{\circ}$ at 150 MHz showing the effect of intrinsic RM depolarisation. The upper panel shows the $2L/3$, $\mathcal{M}=3$ and $\theta_{bn} = 0^{\circ}$ case and the lower panel shows the $L/4$, $\mathcal{M}=3$ and $\theta_{bn} = 0^{\circ}$ case. The gray contours show the surface brightness.}
    \label{fig:polRM_maps}
\end{figure}
%

\subsection{The effect of internal Faraday Rotation}
\label{sec:RM}

\begin{figure*}
    \centering
    \includegraphics[width=0.325\textwidth]{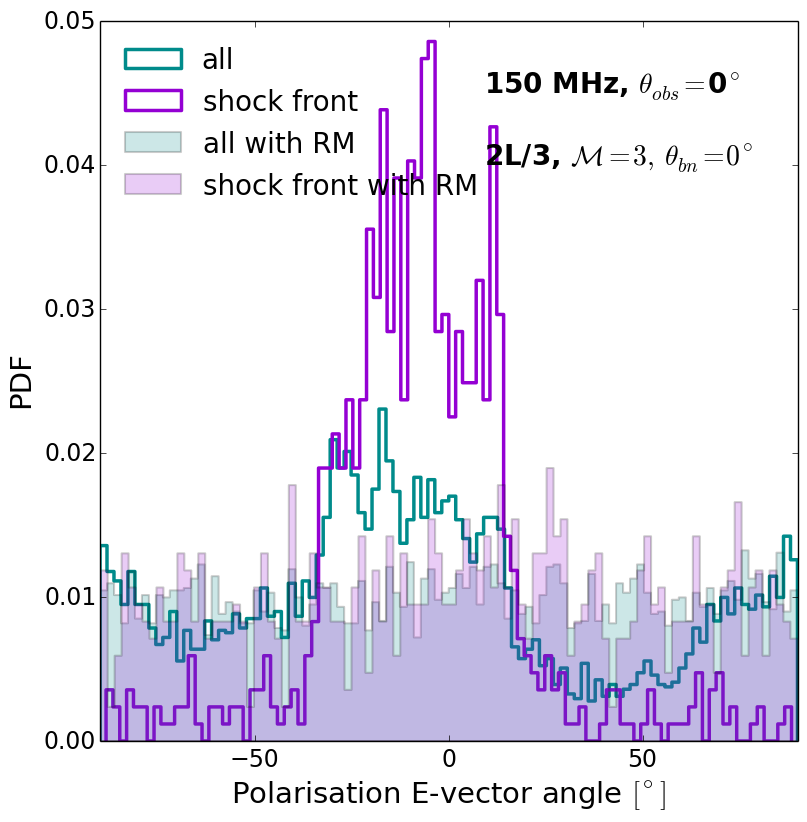}
    \includegraphics[width=0.325\textwidth]{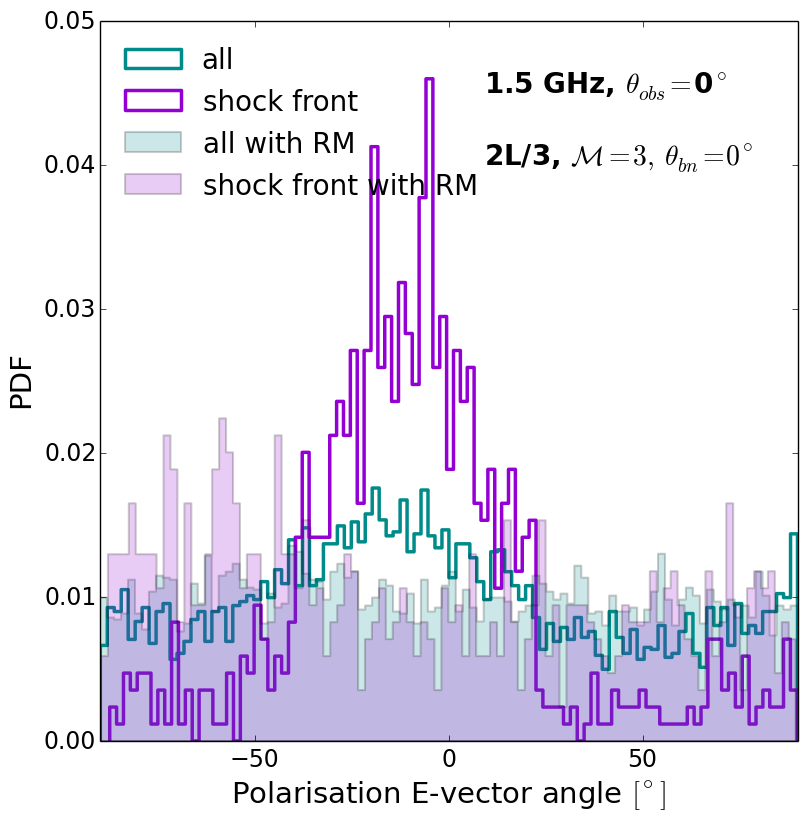}
    \includegraphics[width=0.325\textwidth]{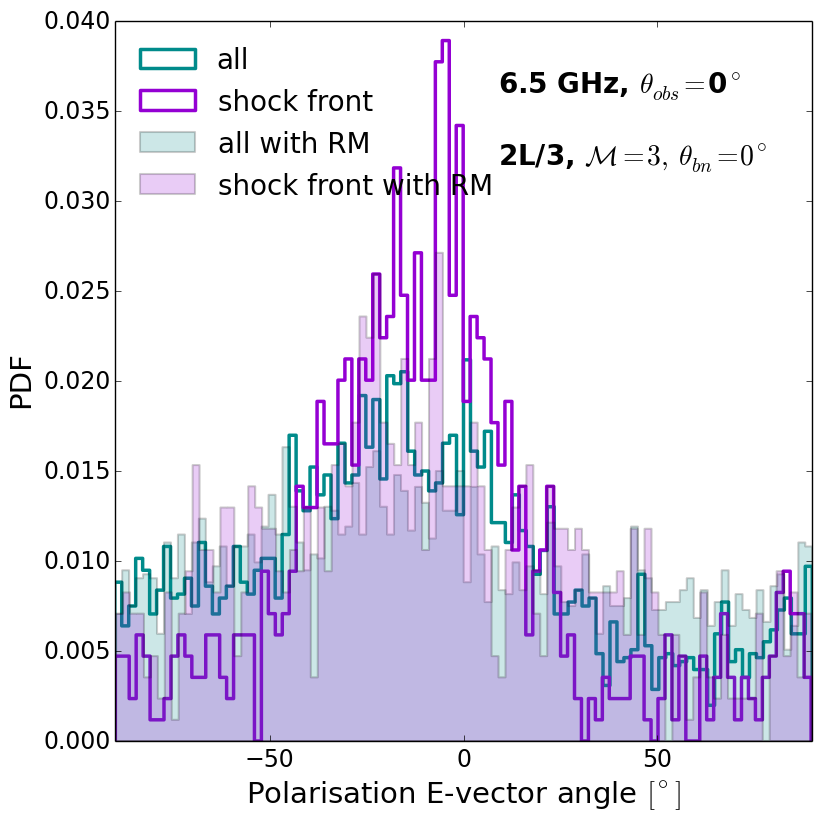} \\
    \includegraphics[width=0.325\textwidth]{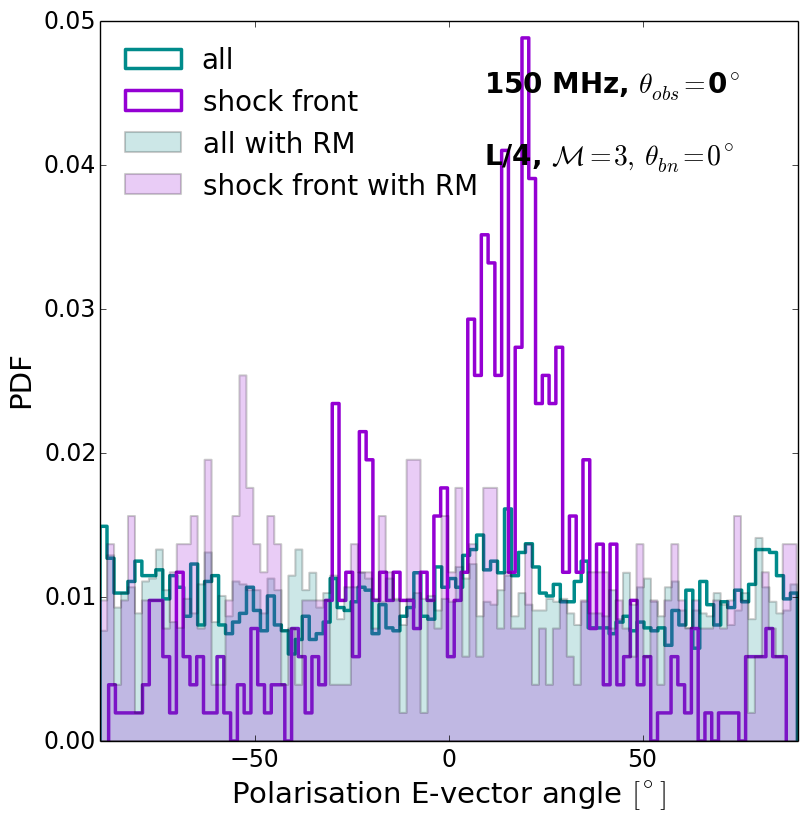}
    \includegraphics[width=0.325\textwidth]{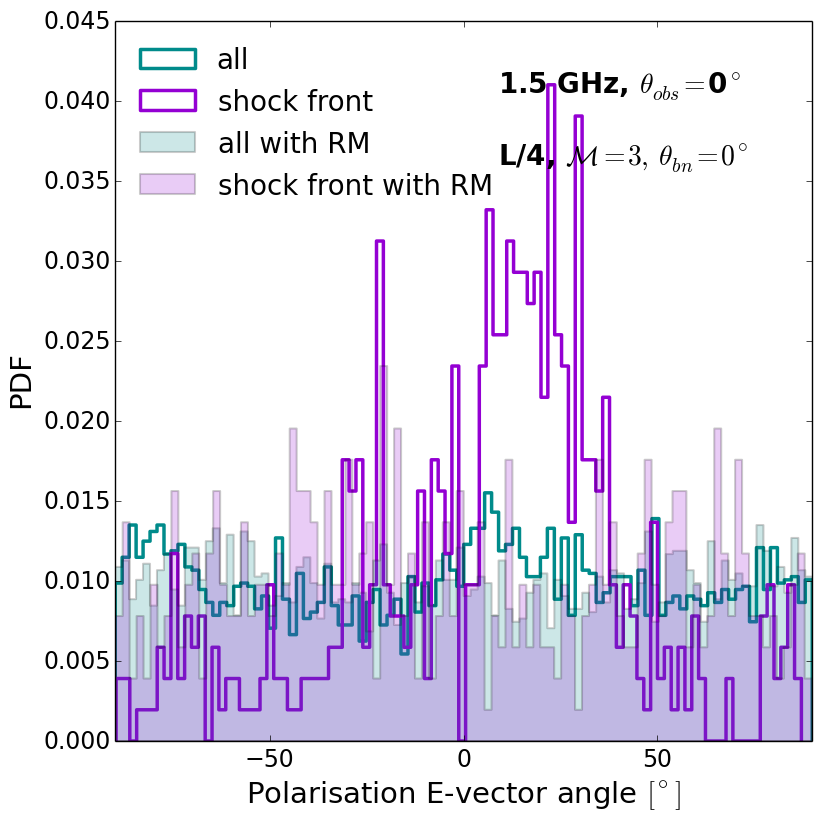}
    \includegraphics[width=0.325\textwidth]{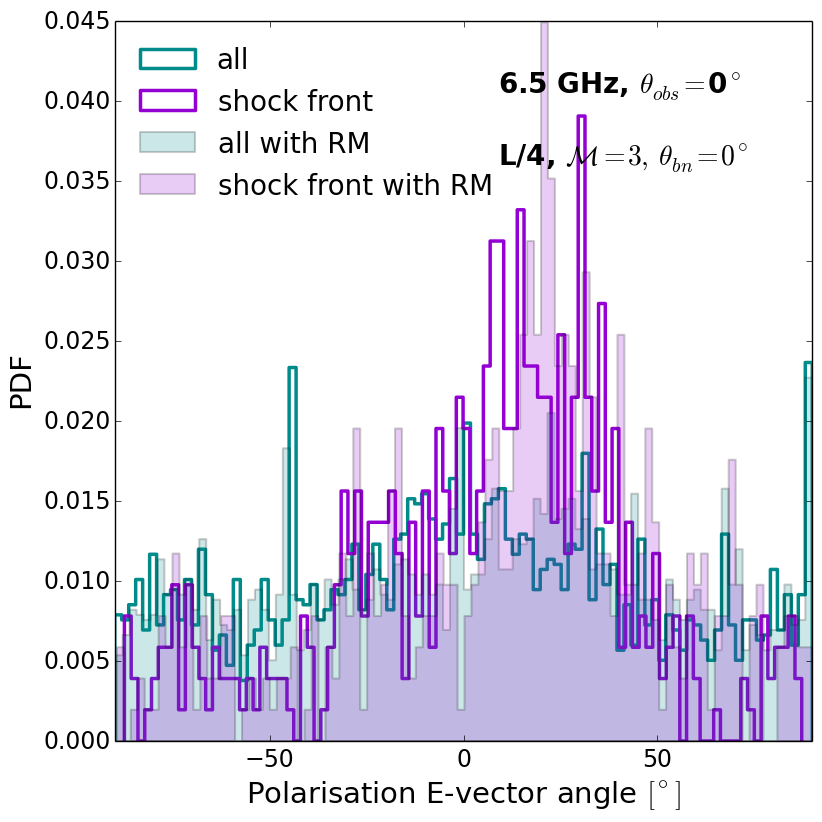} 
    \caption{PDFs of the polarisation $E$--vector angle with (step) and without (step-filled) the RM contribution. We show $\theta_{\rm obs}=0^{\circ}$ at different frequencies. The first row shows the $2L/3$, $\mathcal{M}=3$ and $\theta_{bn} = 0^{\circ}$ case, and the second row shows the $L/4$, $\mathcal{M}=3$ and $\theta_{bn} = 0^{\circ}$ case. The first, second and third columns correspond to the frequencies 150 MHz, 1.5 GHz and 6.5 GHz. We show the PDF of $\sim 60$ kpc in the downstream region in darkcyan and that of the shock front in darkorchid.}
    \label{fig:hist_RM}
\end{figure*}

There are two contributions to consider when the polarised signal propagates through a magnetised medium along the LOS: internal RM (intrinsic to the source) and external RM (external to the source). The external contributions to the observed RM can come from the foreground intergalactic medium, e.g. the contribution from the Milky Way, and from the foreground ICM in the case that the relic is located behind the cluster. 
The local Galactic foreground from the Milky Way can in general be calibrated with RM grid data from our Galaxy \citep[e.g.][]{2020A&A...633A.150H}. On the other hand, the contribution of a larger foreground ICM to the external Faraday rotation is subject to the location of the relic within the galaxy cluster. Observationally, the external contribution can be constrained using RM synthesis and QU-fitting techniques \citep[e.g.][]{2018ApJ...855...41A,2019MNRAS.489.3905S,2020ascl.soft05003P,2020arXiv200906554L}. Theoretically, it can also be studied with cosmological simulations \citep[e.g.][]{2018MNRAS.480.3907V,wittor2019}. Our simulations are dedicated to study a small region of the ICM, 400 kpc $\times$ 200 kpc $\times$ 200 kpc (see Sec.~\ref{section:num_set-up2}), and therefore can only yield the internal contribution. 
In this section, we focus on the Faraday rotation contributed by the ICM within our computational box (e.g. internal RM).

In Fig.~\ref{fig:PDF_RM} we show the PDF of the RM for both turbulent media: at $t=188$ Myr for the $2L/3$ case (upper panel) and at $t=292$ Myr for the $L/4$ case (middle panel). 
We show the PDF for the 3D distribution of RM (measured within single cells) and the 2D distribution obtained after integrating along the full LOS  in the $z$-direction. We note how the RM strongly depends on the density and magnetic field fluctuations in each of our turbulent media, and also how the distribution of the RM evolves in time as the shock is propagating through the ICM. In general, the compression of the turbulent magnetic field increases the RM over time.
For the 2L/3 turbulence model, we obtain RM values ranging from $\pm 200$ rad m$^{-2}$ at the early stages of the shock propagation, up to $\pm 1000$ rad m$^{-2}$ at the late stages of the shock propagation (see 3D distribution in the upper panel of Fig.~\ref{fig:PDF_RM}). In this case, we can also see that the RM distribution at the late stages is non-Gaussian. This non-Gaussianity has also been observed in cosmological simulations \citep[see e.g.][]{wittor2019}. For the L/4 turbulence model, we obtain RM values ranging from $\pm 100$ rad m$^{-2}$ at the early stages of the shock propagation to $\pm 400$ rad m$^{-2}$ at the late stages of the shock propagation (see 3D distribution in the middle panel of Fig.~\ref{fig:PDF_RM}). At the snapshot shown in the upper and middle panels of Fig.~\ref{fig:PDF_RM}, the 2L/3 case has a mean RM value of $\sim -6$ rad m$^{-2}$ and a standard deviation of the RM of about $\sigma_{RM}=194$ rad m$^{-2}$, whereas the L/4 case has a mean RM value of $\sim 1$ rad m$^{-2}$ and a standard deviation of the RM of about $\sigma_{RM}=80$ rad m$^{-2}$. In observations, the RM can only be inferred for regions that show radio emission. Therefore, we also compute the PDF of the RM weighted by the radio emission regions:
\begin{equation}\label{eq:RM_weight1}
    \mathrm{RM}= 0.812 \, \, \mathrm{rad} \, m^{-2}  \sum_{i=0}^{n} n_e B_{\parallel} w_i \, dl,
\end{equation}
where $n=n_z$ (total number of cells in the $z$-axis) for the 2D distribution and $n \in [1,n_z]$ for the 3D distribution. The weight function $w_i$ is defined as
\begin{equation}\label{eq:RM_weight2}
    w_i = \left\{
	     \begin{array}{ll}
		 1       & \mathrm{if\ } \Lambda \neq 0, \\
		 & \\
		 0 & \Lambda = 0, \\
		 
	       \end{array}
	     \right.
\end{equation}
where 
$\Lambda = \mathcal{J}_{\nu}$ (emissivity). 
We selected the emission at 150 MHz as a weight for Fig.~\ref{fig:PDF_RM}.

The PDF of the RM (both in 2D and 3D) narrows when taking into account only the radio emitting cells. For the 2L/3 turbulence, we obtain RM values of $\pm 500$ rad m$^{-2}$ in the 3D distribution at $t=188$ Myr (see the dashed blue line PDF in Fig.~\ref{fig:PDF_RM}). For the L/4 turbulence, we obtain RM values of $\pm 200$ rad m$^{-2}$ in the 3D distribution at $t=292$ Myr (see the dashed green line PDF in Fig.~\ref{fig:PDF_RM}). Note that we have arbitrarily selected a weighting such that the RM reflects radio regions with emissivities larger than the zero. Nonetheless, it is straightforward to see that the PDF could narrow even more if we were to select e.g. the brightest radio cells. Keeping this in mind, these values are in the ballpark of typically observed RM at the outskirts of clusters \citep[e.g.][for a catalog of RM of extragalactic polarised sources]{2016A&A...596A..22B} or for individual sources \citep[see e.g.][]{2004astro.ph.11045J,2013MNRAS.433.3208B,2020A&A...642L..13R,2021arXiv210206631D}.

 In \citetalias{dominguezfernandez2020morphology} we showed that the amplification of the magnetic field proceeds in a different manner for the $2L/3$ and $L/4$ cases. In the same way, we show how the standard deviation of the RM, $\sigma_{\mathrm{RM}}$ changes in time for both models in the lower panel of Fig.~\ref{fig:PDF_RM}. As can be seen, the RM changes can differ considerably depending on the initial ICM conditions. The $\sigma_{\mathrm{RM}}$ will increase as the shock continues to propagate through each medium. In particular, the $\sigma_{\mathrm{RM}}$ increases more for the 2L/3 case than for the L/4 case. This is line with the increment of the standard deviation of the magnetic field (see Fig. 7 in \citetalias{dominguezfernandez2020morphology}). Since the rate of change will depend on the initial magnetic field strength, the rms Mach number in the medium, energy ratios, the characteristic scale of the underlying fluctuations and the strength of the shock, it is not possible yet to fully parametrise its evolution as a function of environmental parameters.

Next we compute the RM contribution to the polarisation in post-processing. The polarisation emissivity can be written as
\begin{equation}\label{eq:pol_withRM}
    \mathcal{P}_{\nu} = \mathcal{J}_{\mathrm{pol}}(\nu_{\mathrm{obs}},x,y,z) \exp{\left[ 2i(\hat{\chi} + \mathrm{RM}\lambda_{\mathrm{obs}}^2) \right]},
\end{equation}
where we have used Eq.~(\ref{eq:RM_angle}). It is then easy to see that the real part of Eq.~(\ref{eq:pol_withRM}) gives the Stokes Q parameter and the imaginary part gives the Stokes U parameter, equivalent to Eqs.~(\ref{Q_stokes}--\ref{U_stokes}). After some algebra we can finally obtain the Stokes Q and U maps with the RM contribution as:
\begin{align}\label{Q_stokes_RM}
\begin{split}
    Q_{\nu} = \int \mathcal{J}_{\rm pol} [ & \cos{(2\hat{\chi})} \cos{(2\mathrm{RM} \lambda_{\mathrm{obs}}^2)} \, - \\ 
    & \sin{(2\hat{\chi})}\sin{(2\mathrm{RM} \lambda_{\mathrm{obs}}^2)}
    ] dz,
\end{split}
\end{align}
\begin{align}\label{U_stokes_RM}
\begin{split}
    U_{\nu} = \int \mathcal{J}_{\rm pol} [ & \sin{(2\hat{\chi})} \cos{(2\mathrm{RM} \lambda_{\mathrm{obs}}^2)} \, + \\ 
    & \cos{(2\hat{\chi})}\sin{(2\mathrm{RM} \lambda_{\mathrm{obs}}^2)} ] dz.
\end{split}
\end{align}

The polarisation fraction with the RM contribution is finally computed in the same way as in Eq.~(\ref{eq:pol_degree}).

We start by showing how the polarisation fraction evolves at the location of the shock front at different frequencies. 
In Fig.~\ref{fig:Polfrac_RM} we show the mean polarisation fraction as function of time for both turbulent media. The mean was computed by taking only into account the polarisation fraction at the shocked cells (similar to Sec.~\ref{sec:pol_uniform}).
We also show how the mean polarisation fraction decreases when the Faraday rotation is taken into account for different frequencies (see dashed lines in Fig.~\ref{fig:Polfrac_RM}). We find that at 150 MHz the RM contribution depolarises both media down to $\sim 10$--20\%. Indeed, observed radio relics are not expected to show significant polarisation at such low frequencies  \citep{2008A&A...489...69B,2011A&A...525A.104P,2015PASJ...67..110O}. At higher frequencies, the Faraday rotation effect becomes less significant. The RM contribution at 6.5 GHz is negligible leading to minor or almost no changes in the evolution of the mean polarisation fraction. On the other hand, the RM contribution at 1.5 GHz depolarises more the $2L/3$ case than the $L/4$ case due to the larger RM values in the former case (see Fig.~\ref{fig:PDF_RM}).

The effect of Faraday depolarisation at 150 MHz can be better visualised in Fig.~\ref{fig:polRM_maps} where we show the polarisation E-vectors with the RM contribution for both turbulent media. We also show the corresponding polarisation E-vectors without RM as previously shown in Fig.~\ref{fig:pol_maps_2turb}. The RM introduces extra rotation in the polarisation E-vectors and it depolarises the downstream and shock front regions, as expected. As shown in Fig.~\ref{fig:Polfrac_RM}, the mean polarisation fraction at the shock front reaches $\sim 10$\% for the L/4 case, while it reaches $\sim20$\% for the 2L/3 case. Note that these values are derived from the $\theta_{\mathrm{obs}}=0^{\circ}$, i.e. edge-on view of the shock. For other viewing angles where $\theta_{\mathrm{obs}}\lesssim 90^{\circ}$, we expect a lower, yet not null, polarisation fraction. The multi-scale structure of a turbulent magnetic field can allow some components to contribute to the linearly polarised emission along the LOS (see Sec.~\ref{sec:pol_methods}). Since it is non-trivial to pin-point the relation between expected polarisation trends and the contributing magnetic components along different LOS, we leave a complete study on viewing angles for future work.

We computed the PDF of the polarisation E-vectors for both turbulent media, with and without the RM contribution. We show the PDFs in Fig.~\ref{fig:hist_RM} for the three different frequencies considered in this work. We additionally show the PDF corresponding only to the cells tracking the shock front (same as in Fig.~\ref{fig:align_150}). It becomes then more evident that the alignment or orientation of the polarisation E-vectors with the shock normal will persist at 6.5 GHz, regardless of the properties of the turbulent medium. On the other hand, already at  1.5 GHz, the RM contribution can be relevant depending on the type of underlying turbulence. These differences are again due to the magnetic power and to the standard deviation of the magnetic field as discussed in Sec.~\ref{sec:pol_turb}. In \citetalias{dominguezfernandez2020morphology}, we showed that a $\mathcal{M}=3$ shock induces a larger increase in the standard deviation of the magnetic field in the 2L/3 case,  compared to the L/4 case. This can be also seen in the lower panel of Fig.~\ref{fig:PDF_RM} where we show the evolution of the RM standard deviation of both cases.

  From the results in this section we can already confirm that in the presence of turbulence, the impact of internal RM on the observable polarised emission is significant at LOFAR-alike or lower frequencies. Its contribution will effectively depolarise the source, likely explaining the typically reported lack of (or little) polarised emission in low-frequency observations of radio relics. Conversely, our study shows that the alignment of the polarisation E-vectors with the shock normal and a high polarisation fraction at the shock front are results that should persist with or without the contribution of internal RM at about 6.5 GHz, or larger. Finally, we find that these results are more dependent on the turbulent conditions when observed at 1.5 GHz. We expect that the exact relation between the observed changes in the polarisation properties and the turbulence parameters might be constrained with future, large scan of model parameters prescribing the level of turbulence in the ICM. 

As a side note, we note that further depolarisation can occur due to instrumental effects such as bandwidth depolarisation. The polarisation angle rotates over the bandwidth effectively depolarising the signal \citep[see][]{2015gimf.book.....K}. In order to minimise this effect, the observing bandwidth is subdivided into many sub-bands or channels and applying RM synthesis \citep[][]{2005A&A...441.1217B}. In the case of LOFAR, a typical observing sub-band is 0.2 MHz and there are 4 to 16 channels in the sub-band \citep[][]{2013A&A...556A...2V}. Assuming a typical observation with 4 channels, the channel width is $\sim$0.05 MHz. At 1.5 GHz the typical channels widths are $\sim$1 MHz and at 6.5 GHz $\sim$10 MHz. The bandwidth depolarisation is given by \citep[see Chapter 3][]{2015gimf.book.....K} 
\begin{equation}\label{eq:bandwidth}
    \Pi_{\nu,b.d} = \Pi_{\nu} \frac{\sin{\Delta \Psi}}{\Delta \Psi},
\end{equation}

where b.d stands for bandwidth depolarisation, $\Pi_{\nu}$ is the polarisation fraction defined in Eq.~(\ref{eq:pol_degree}) and 
\begin{equation}\label{eq:extra}
    \Delta \Psi = - 2 \lambda_0^2 RM \, \frac{\Delta \nu}{\nu_0},
\end{equation}

where $\lambda_0$ and $\nu_0$ are the observing central wavelength and frequency, respectively. In Table \ref{table:bandwidth_depol} we show the amount of bandwidth depolarisation expected at 150 MHz, 1.5 GHz and 6.5 GHz for the two turbulent models. Note that here we compute $\Delta \Psi$ assuming the maximum RM to be conservative (see Figure \ref{fig:PDF_RM}). This will make a small difference at 150 MHz in the polarisation fraction only in the 2L/3 model. Nevertheless, if we take the mean RM to be the representative value in Eq.~(\ref{eq:extra}), the polarisation fraction will not change. Since the maximum RM values are only achieved at certain cells, it is overall safe to say that the effect of bandwidth depolarisation can be ignored in this work. 

\begin{table}
\centering
\begin{tabular}{cccc}
    & \\ \hline
      $\nu_0$ [GHz] & $\Delta \nu$ [MHz] &  $\frac{\sin{\Delta \Psi}}{\Delta \Psi}\vert_{2L/3}$ & $\frac{\sin{\Delta \Psi}}{\Delta \Psi}\vert_{L/4}$ \\ \hline
      0.15 &  0.05  & 0.73 & 0.95 \\ 
      1.5 &  1.0  & 0.99 & 1.0 \\ 
      6.5 &  10 & 1.0 & 1.0 \\ 
      \hline
\end{tabular}
\caption{First column: central frequency studied in this work. Second column: typical channel width from observations at the selected frequencies. Third column: bandwidth depolarisation term in Eq.~(\ref{eq:bandwidth}) using the maximum RM value showed in the weighted PDF in Fig.~\ref{fig:PDF_RM} for the 2L/3 case ($\sim 500$ rad m$^{-2}$). Fourth column: same as third column but for the L/4 case ($\sim 200$ rad m$^{-2}$).}
\label{table:bandwidth_depol}
\end{table}

\begin{figure}
    \centering

    \includegraphics[width=0.48\columnwidth]{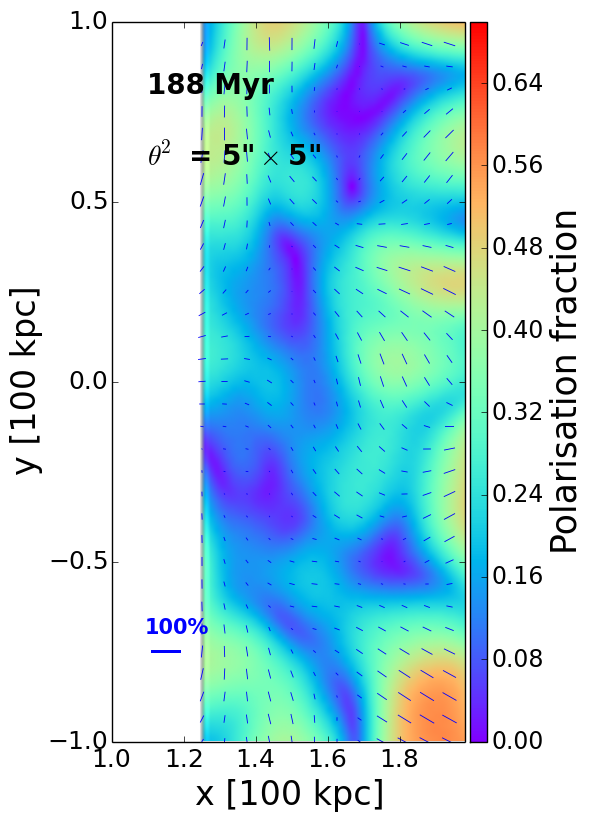}
    \includegraphics[width=0.48\columnwidth]{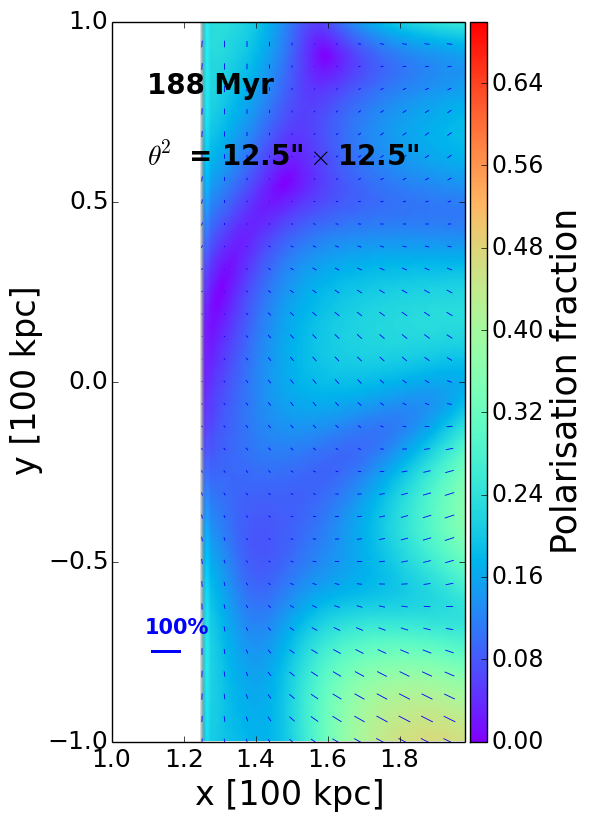} \\
    \includegraphics[width=0.48\columnwidth]{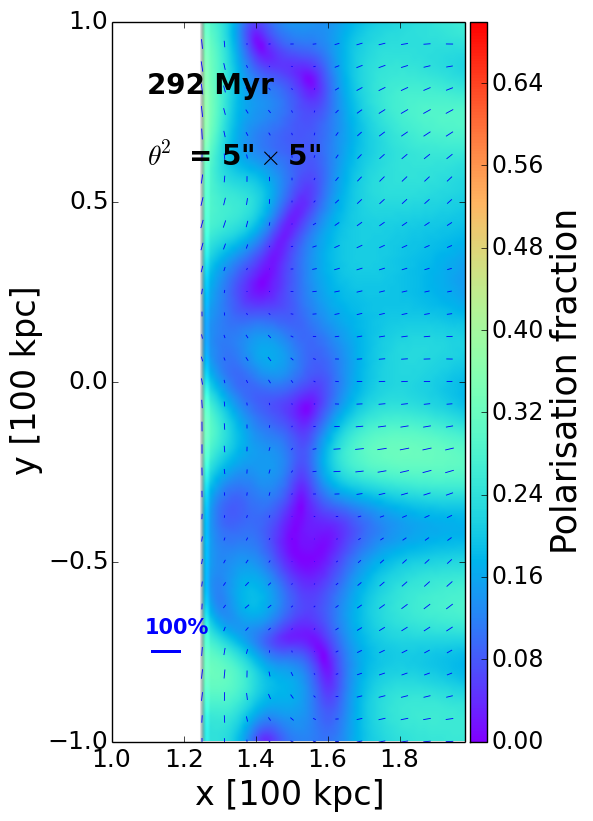}
    \includegraphics[width=0.48\columnwidth]{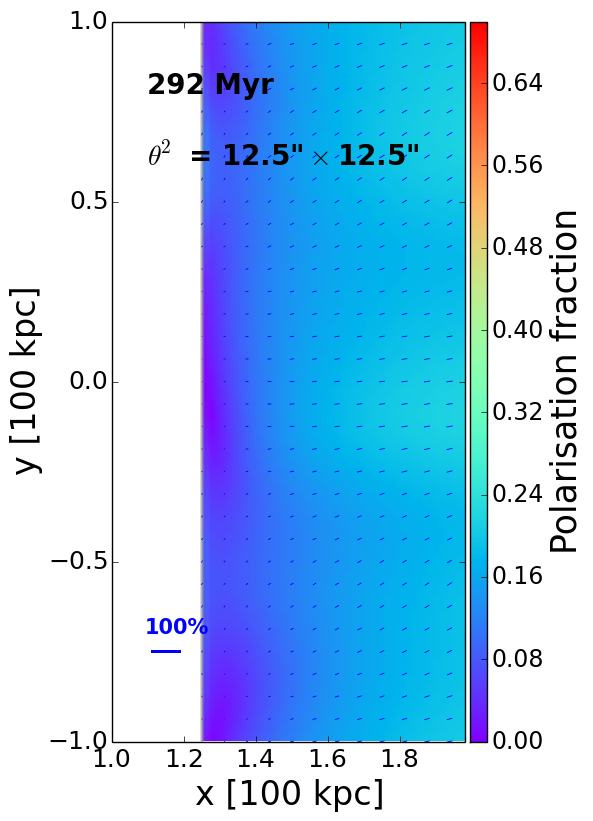} 
    \caption{Polarisation fraction maps smoothed considering $\theta_{\rm obs}=0^{\circ}$ at 6.5 GHz. The first row shows the $2L/3$, $\mathcal{M}=3$, and $\theta_{bn} = 0^{\circ}$ case and the second row shows the $L/4$, $\mathcal{M}=3$, and $\theta_{bn} = 0^{\circ}$ case. The first and second columns correspond to beam sizes of 5"$\times$5" and 12.5"$\times$12.5", respectively. We overplot the polarisation $E$-vectors obtained with the smoothed Q and U maps and Eq.~(\ref{eq:pol_angle}). Note that here we have included the internal RM contribution.}
    \label{fig:beam_maps}
\end{figure}

As a final note, turbulence may extend to smaller scales than our current resolution. Hence, in reality, we would find at each cell a distribution of magnetic field vectors which may lead to extra depolarisation at each cell. In order to asses this possible outcome, we can assume as a first-order approximation that the slope for the magnetic power spectrum is the same as the Kolmogorov $k^{-5/3}$ spectrum. In this case, let us assume that in a cell the magnetic field fluctuates on scales that are a fraction of our resolution, i.e.  $f\Delta x$  with $0 \leq f\leq 1$. For $f=0.1$, we would obtain a fluctuations of about $\sim 0.1^{5/3}\sim 0.02$, and for $f=0.3$ of about $\sim 0.3^{5/3}\sim 0.13$. \citealt{1998MNRAS.299..189S} estimated the depolarisation fraction due to an isotropic random magnetic field superimposed on a
regular magnetic field. We considered their Eq. (24),

\begin{equation}
    \Pi_{\nu,S} = \, \Pi_{\nu} \frac{\overline{B}_{\perp}^2}{\overline{B}_{\perp}^2 + 2\sigma_B^2} = \, \Pi_{\nu}\, W_S, 
\end{equation}
where $\Pi_{\nu,S}$ denotes the polarisation fraction in Sokoloff's work, $\sigma_B$ is the one-dimensional standard deviation of the random magnetic field and $\overline{B}_{\perp}$ is the mean magnetic field perpendicular to the LOS. The $W_S$ value from the turbulent random magnetic field is estimated to be $\sim$ 0.92--0.99 considering $f=0.1$ and $f=0.3$\footnote{Note that the 3D components are related to the 1D component through $\sigma_x^2 + \sigma_y^2 + \sigma_y^2 = 3\sigma_B^2$}. Therefore, the intrinsic polarisation could be lower by a $\sim$ 0.92--0.99 due to the tangled field on unresolved scales. It shall be noted that compression of a merger shock may also cause the unresolved magnetic field distribution to be anisotropic, hence the depolarisation could be lower than estimated above. In future work, we will asses this by analysing higher resolution simulations.

\subsection{Beam effects and depolarisation}
\label{sec:beam}

\begin{figure}
    \centering
    \includegraphics[width=0.32\textwidth]{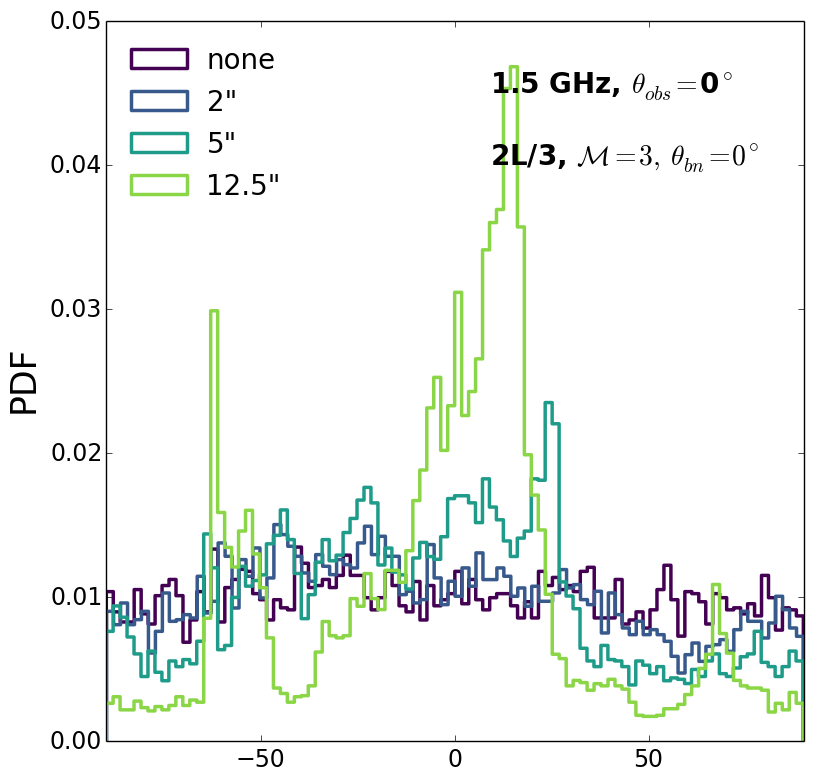}
    \includegraphics[width=0.32\textwidth]{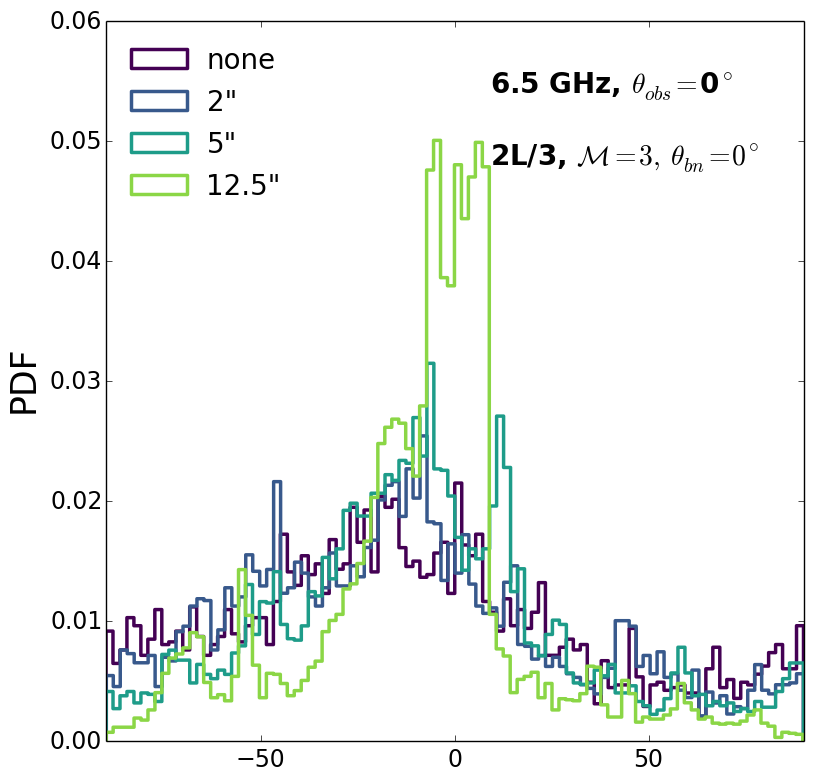} \\
    \includegraphics[width=0.32\textwidth]{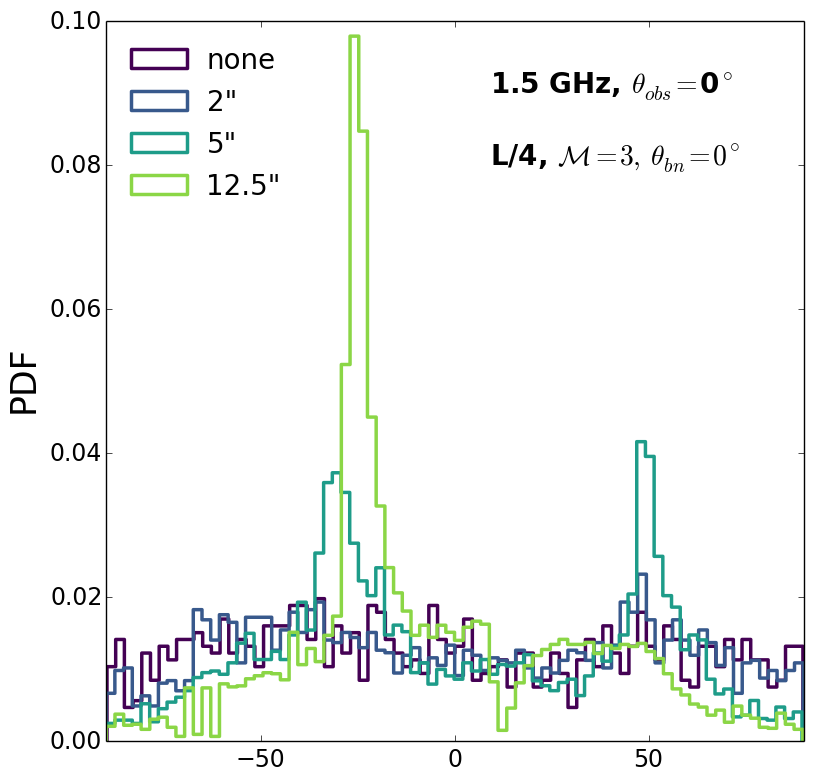}
    \includegraphics[width=0.32\textwidth]{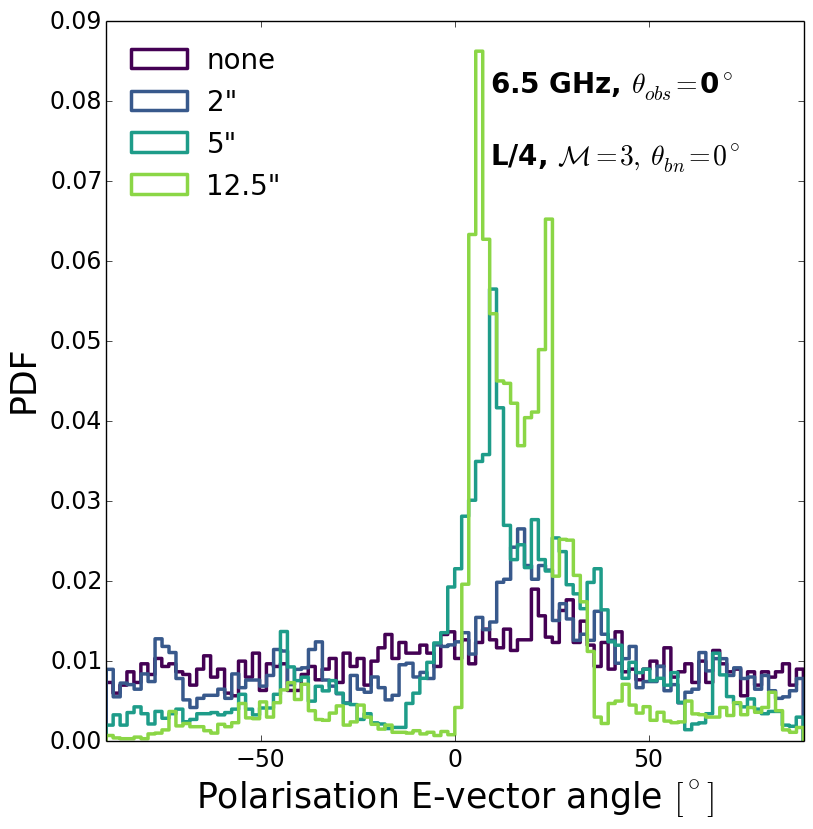} 
    \caption{PDFs of the polarisation $E$--vector angles considering $\theta_{\rm obs}=0^{\circ}$ at different frequencies and different beam sizes (see Table \ref{table:beams}). The two upper panels show the $2L/3$, $\mathcal{M}=3$ and $\theta_{bn} = 0^{\circ}$ case, and the two lower panels show the $L/4$, $\mathcal{M}=3$ and $\theta_{bn} = 0^{\circ}$ case. The first (second) and third (fourth) panels correspond to 1.5 GHz (6.5 GHz).}
    \label{fig:beam_hist}
\end{figure}

The observed polarised emission is inevitably tied to a beam size. The selected beam size can change the observed polarisation fraction, as well as the observed orientation of the polarisation $E$- and $B$-vectors. This effect is known as beam depolarisation, and could be stronger when the scale of magnetic fluctuations is smaller than the beam size. In this section we study how different beam sizes change the results from Sec.~\ref{sec:RM}. We consider different beam sizes in order to smooth the $I_{\nu}$, $Q_{\nu}$, and $U_{\nu}$ maps. We assume a redshift $z=0.22$ corresponding to that of the merging galaxy cluster 1RXS J0603.3+4214 which hosts the Toothbrush relic. We show the beam characteristics in Table~\ref{table:beams}. In Sec.~\ref{sec:RM} we showed how internal Faraday rotation depolarises the emission mainly at 150 MHz. Therefore, we focus on discussing beam depolarisation at 1.5 GHz and 6.5 GHz in this section.

In Fig.~\ref{fig:beam_maps} we show the polarisation fraction maps 
at 6.5 GHz
smoothed on the scales of two different beams: 5"$\times$5" and 12.5"$\times$12.5". These maps were obtained by first applying Gaussian smoothing with a kernel corresponding to the desired FWHM in the $I_{\nu}$, $Q_{\nu}$, and $U_{\nu}$ maps. In this case, we have used Eqs.~(\ref{Q_stokes_RM}--\ref{U_stokes_RM}) where the Faraday rotation effect is taken into account in order to show a more realistic scenario. In the first row, we show the $2L/3$, $\mathcal{M}=3$, and $\theta_{bn} = 0^{\circ}$ case and the second row, the $L/4$, $\mathcal{M}=3$, and $\theta_{bn} = 0^{\circ}$ case. The lower the resolution, the lower the polarisation fraction, i.e. we observe beam depolarisation when going to larger beams/lower resolutions.

The decreasing amplitude of fluctuations in the polarisation fraction is also evident at lower resolutions for both turbulent length scales. 
For example, at the lowest resolution, i.e. the beam of 12.5"$\times$12.5", 
the local polarisation fraction reaches up to 
$\sim$45\% and $\sim$20\% at the shock front in the $2L/3$ and $L/4$ cases, respectively.

Having a higher polarisation fraction at regions of the shock front and depolarisation in the downstream even at lower resolution seems to be a pattern that cannot be reproduced with a uniform medium.
We saw in Sec.~\ref{sec:pol_uniform} that a uniform medium produces a gradient with the highest values at the downstream region and the lowest at the shock front (see Fig.~\ref{fig:pol-uni2}). This gradient in the uniform runs remains observable when considering the three different beam sizes. We show this result in Fig.~\ref{fig:beam_uni} of Appendix \ref{sec:app1}. 
Therefore, Fig.~\ref{fig:beam_maps} shows that even at low resolution, we would distinguish between a turbulent or uniform media at the relic position. We also can confirm that a turbulent medium with smaller magnetic fluctuations is more affected by beam depolarisation. Overall, Fig.~\ref{fig:beam_maps} shows drastic morphological changes in the observed polarised emission due to a limited resolution. 

\begin{table}
\centering
\begin{tabular}{cc}
    & \\ \hline
      Beam & FWHM [kpc]   \\ \hline
      2"$\times$2" &  7.4  \\ 
      5"$\times$5" &  18.4  \\ 
      12.5"$\times$12.5" &  46.1  \\ 
      \hline
\end{tabular}
\caption{Restoring beam sizes typical of the VLA (in either configuration BCD or ABCD) assumed for smoothing the emission maps. We considered a redshift of $z=0.22$ to quote the corresponding linear length of the full width half maximum (FWHM).}
\label{table:beams}
\end{table}

The PDF of the polarisation $E$-vector angles at 1.5 and 6.5 GHz 
can be observed in Fig.~\ref{fig:beam_hist}. 
At higher frequencies, the $2L/3$ case shows a stronger alignment of the polarisation $E$-vectors with the $x$--axis (which roughly corresponds to the direction of the shock's normal vector). This can be seen in the shape of the PDF with no smoothing, whereas the PDF of the $L/4$ case is in general flatter than the PDF for the $2L/3$ case, confirming the more random orientation of polarisation vectors. Note that here we are considering the emission from the entire downstream region and not just the shock front as in Fig.~\ref{fig:align_150}. An angular resolution of 2" leaves the shape of the PDF roughly unchanged in all runs (see Fig.~\ref{fig:hist_RM}). However, poorer resolutions change significantly the orientation of the polarisation $E$--vectors. In particular, a 12.5" resolution tend to result in smaller angles of the $E$--vectors (defined with respect of our $x$--axis), i.e narrowing the PDF.  
The reason is that larger beams give more weight to the brightest regions where the $E$-vector is largely aligned with the shock normal, i.e $\psi \sim 0^{\circ}$ as shown in Fig.~\ref{fig:align_150} (see also Fig.~\ref{fig:evol} and Fig.~\ref{fig:profiles_all}).

This effect is evident at 6.5 GHz, and it becomes more subtle at 1.5 GHz. As discussed in Sec.~\ref{sec:RM}, the RM contribution at 1.5 GHz can be significant and it depends on the turbulent medium. 
Depolarisation could lead to a flatter PDF of the polarisation $E$-vector angle, with no particular preference or peak at a certain angle. But even in this case, a lower resolution image will
tend to pick up the brightest regions and to narrow the range of the observed PDF. Therefore, a lower resolution image can be biased compared to the intrinsic PDF, as shown in Fig.~\ref{fig:beam_hist}.

\begin{figure}
    \centering
    \includegraphics[width=0.9\columnwidth]{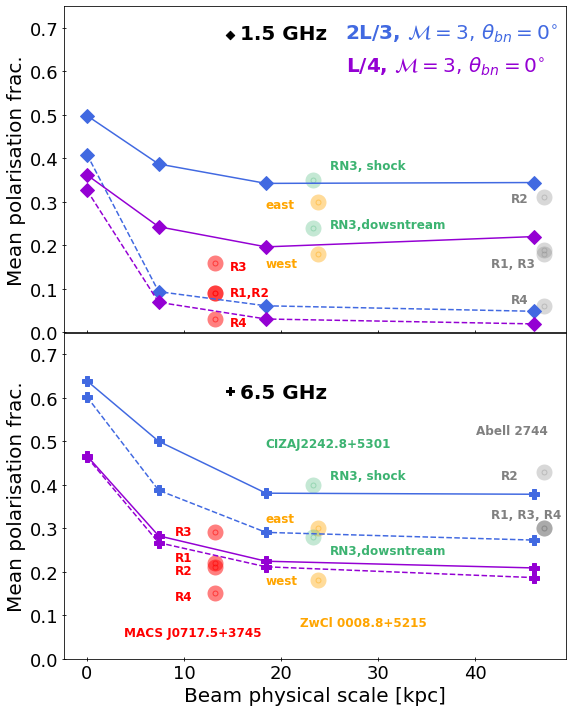}
    \caption{Mean polarisation fraction at 1.5 GHz (\textit{upper panel}) and 6.5 GHz (\textit{lower panel}) of the of $\sim 60$ kpc downstream region versus resolution (see Table \ref{table:beams}) for our two models. We show the mean polarisation fraction without the contribution of RM (\textit{solid lines}) and with the RM contribution (\textit{dashed lines}). We overplot data from the regions of the relic: MACS J0717.5+3745 (red: 1.5 GHz (\textit{upper panel}) and 5.5 GHz (\textit{lower panel}); {\color{blue} Rajpurohit et al. submitted})), Abell 2744 (gray: 1.5 GHz (\textit{upper panel}; \citealt{rajpurohit2021dissecting}) and 2--4 GHz (lower panel;  R1 from \citealt{rajpurohit2021dissecting} and R2, R3 and R4 from \citealt{2017ApJ...845...81P})), ZwCl 0008.8+5215 (orange: 2--4 GHz (same data in both panels); \citealt{2017ApJ...838..110G}) and CIZAJ2242.8+5301 (green: 1.5 GHz average values at the shock location and at the downstream (\textit{upper panel}) and best-fit average values at the corresponding regions (\textit{lower panel}); \citealt{2021arXiv210206631D}). 
    }
    \label{fig:obs}
\end{figure}

Finally, in Fig.~\ref{fig:obs}, we summarise how the mean polarisation fraction changes with increasing beam sizes 
at 1.5 GHz and at 6.5 GHz. We show the mean polarisation of the $\sim$60 kpc downstream region with and without the RM contribution (see Sec.~\ref{sec:RM}).
For the redshift considered here ($z=0.22$) we measure a significant drop of polarisation for beam resolutions below 2". 
This gradual decline will clearly depend not only on the size of the beam, but also on the underlying turbulent medium. The 2L/3 case decreases from $\sim$50\% to $\sim$40\% at 1.5 GHz and from $\sim$65\% to $\sim$45\% at 6.5 GHz. On the other hand, the L/4 case decreases from $\sim$40\% to $\sim$30\% at 1.5 GHz and from $\sim$50\% to $\sim$25\% at 6.5 GHz. These differences between both models are tightly related to 1) the amount of compression at the shock front and 2) how the $B_x$-$B_y$ distribution becomes more isotropic towards the downstream (see discussion in Sec.~\ref{sec:pol_turb}).  

The effect of beam depolarisation becomes more significant if we consider internal Faraday rotation (see dashed lines in Fig.~\ref{fig:obs}). At 1.5 GHz, the mean polarisation fraction of both models can decrease to $\lesssim1$\% at the lowest resolution. At 6.5 GHz, the mean polarisation fraction can drop to $\sim20$--30\%. In this case, the effect of Faraday rotation is smaller and as a consequence the decrease is more subtle. The 2L/3 case suffers of more Faraday depolarisation than the L/4 case due to the larger RM values (see Sec.~\ref{sec:RM}). This aggravates the depolarisation at lower resolution in the 2L/3 case.

We compare these values with data from three relics, that can qualitatively be compared to our work: 
\begin{itemize}
    \item[1)] The relic in MACS\,J0717.5+3745 \citep[see][]{van_Weeren_2017,2018MNRAS.478.2927B,rajpurohit2020understanding} with a Mach number of $\mathcal{M} = 3.3$, as inferred from the injection spectral index $\alpha_{\rm int}$ in \citet{rajpurohit2020understanding},
exhibits polarisation intensity values in the VLA L-band at 1--2 GHz of 0.01--0.06 mJy/beam with a resolution of 5" (e.g. \citealt{2009A&A...503..707B}, {\color{blue} Rajpurohit et al. submitted}). The authors observe that the polarisation fraction increases with increasing angular resolution, which is in agreement with our results. The low mean polarisation fraction observed in the relic can plausibly be explained by a turbulent ICM if internal Faraday rotation is included in the polarisation computation. We can see that the observation values at 1.5 GHz for all regions of the relic and specially for regions R1 and R2 lie within the expectations of both turbulent models (see upper panel of Fig.~\ref{fig:obs}). At 6.5 GHz, the L/4 case seems to be favoured (see lower panel of Fig.~\ref{fig:obs}). However, the observational values reported in the lower panel of the figure are obtained at 5.5 GHz, in which case we would expect a slightly lower mean polarisation in our data points.

 \item[2)] The polarised relics observed in Abell 2744 at $z=0.308$ \citep{2017ApJ...845...81P} with Mach numbers ranging from $\mathcal{M}\sim 2$ (R2, R3 and R4) to $\mathcal{M}\sim 3$ (R1; see also \citealt{Paul_2019}).  At 2--4 GHz and a resolution of 10", R1 shows a polarisation fractions $\sim$52\% locally and $\sim 27$\% globally. 
 When we compare these observational values to our models at 1.5 GHz, we note that only the results with internal Faraday rotation are in line with
 the observed mean polarisation fraction in R4, whereas R1, R2 and R3, are in line with the results without internal Faraday rotation. On the other hand, the relics R1, R3 and R4 have a mean polarisation fraction that lies within the results of the 2L/3 model with Faraday rotation at 6.5 GHz. Yet, if the data is observed at lower frequencies, i.e. 2--4 GHz, this suggests that the two turbulent models presented in this work yield too low values to explain the mean polarisation fraction in Abell 2744. We note that the observational data showed in the lower panel from \citealt{2017ApJ...845...81P}  is not corrected by the effect of Faraday rotation, so these values may change. As of yet, our results suggest that the majority of relics in Abell 2744 may be better explained by a magnetic configuration with larger characteristic scales than our 2L/3 model and/or with a lower magnetic field strength such that the effect of internal Faraday depolarisation is reduced.
  
  \item[3)] The relics in the cluster ZwCl\,0008.8+5215 at $z=0.104$  \citep[see][]{2017ApJ...838..110G} with Mach numbers ranging from $\mathcal{M}=$2.2 (eastern relic) to $\mathcal{M}=$2.4 (western relic), have polarisation fractions of $\sim$30\% and $\sim$18\%, respectively, at 2-4 GHz and a resolution of 12". 
  The comparison of these observational values to both our turbulent models entails very similar conclusions to those for R1, R2 and R3 in Abell 2744.
  The eastern relic's alignment of the polarisation $E$-vectors and the level of polarisation suggests that this relic could also be an outcome of a mild shock propagating through a turbulent ICM only in the case where the magnetic field is coherent in larger scales than that of our 2L/3 model. 
  
  \item[4)] The northern relic in the galaxy cluster CIZAJ2242.8+5301 at $z=0.1921$ \citep{2010Sci...330..347V,2011MNRAS.418..230V} with a Mach number of $\mathcal{M}=2.58$, as inferred from the injection spectral index in \citealt{2018ApJ...865...24D}. Recently, \citealt{2021arXiv210206631D} did a high-resolution study of the polarisation properties of this relic. The authors report RM values ranging from $\sim -150$ rad m$^{-2}$ to $\sim 130$ rad m$^{-2}$. The intrinsic polarisation fraction 7" resolution maps at 1.5 GHz reveal values of up to $\sim60$\% and down to $\sim20$\%. Comparing with our mean intrinsic polarisation values at 1.5 GHz (solid lines in Fig.~\ref{fig:obs}), we see that the values fall in the expectations for both turbulent models. The observational mean values are reported at the shock front and the downstream, so we expect the mean value of both regions to be in between the former two. This favours no particular model at 1.5 GHz. Nevertheless, we would expect those values to be slightly higher at 6.5 GHz. In this case, the 2L/3 model would be favoured. Overall, the variations in the polarisation fraction together with the first-time observed depolarisation towards the downstream makes this relic a clear example of a mild shock propagating through a turbulent ICM.
\end{itemize}

\section{Summary and conclusions}
\label{sec:conclusions2}

Radio relics have a high polarisation fraction, up to  65\% \citep[e.g.][]{2010Sci...330..347V,2012A&A...546A.124V,2014ApJ...794...24O,2017A&A...600A..18K,2020MNRAS.498.1628L,2020A&A...642L..13R}, with  the electric field vector often aligned with the shock normal \citep[e.g.][]{2010Sci...330..347V,2017ApJ...845...81P,2017ApJ...838..110G}. The cause of the high polarisation fraction and of the alignment of the magnetic field with the shock surface remain unclear. Possible causes could be large-scale, uniform fields or compression of a randomly orientated magnetic field \citep[e.g.][]{1980MNRAS.193..439L}. However, some radio relics show more complex polarisation patterns. In particular, some observed radio relics show high polarisation fractions at the shock front contrary to theoretical expectations (e.g. \citealt{2020A&A...642L..13R}, {\color{blue} Rajpurohit et al. submitted} and \citealt{2021arXiv210206631D}).

We have studied the intrinsic polarised emission from a shock wave propagating through a medium perturbed by decaying subsonic turbulence in the ICM. In our hybrid simulations, the MHD grid represents a thermal fluid, whereas Lagrangian particles represent CR electrons. We injected CR electrons at the shock assuming diffusive shock acceleration, after which each CR electron evolves at run-time according to a cosmic-ray transport equation including adiabatic, synchrotron, and inverse Compton losses.

We explored shocks with $\mathcal{M}=3$, typical of merger shock waves. We varied the downstream turbulence using turbulence-in-a-box simulations as previously considered in \citetalias{dominguezfernandez2020morphology}: solenoidal subsonic turbulence with power peaking at 133 kpc ($2L/3$ case) and  solenoidal subsonic turbulence with power peaking at 50 kpc ($L/4$ case). One snapshot of each simulation was selected as an initial condition for our subsequent shock simulation.  We also varied the observing angles, beam sizes and observing frequencies. Our most important results can be summarised as follows:

\begin{itemize}
    \item[i)] \textit{Intrinsic polarisation fraction}: The morphology of polarised emission resembles the structures of the underlying turbulent ICM. In particular, we observe a high polarisation fraction at the shock front, which decreases in the downstream region. We find that the degree of anisotropy in the downstream magnetic configuration defines the amount of depolarisation behind the shock. Since this trend is also observed in some of the best-studied relics such as the Sausage relic \citep[e.g.][]{2021arXiv210206631D} or that in the MACSJ0717.5+3745 galaxy cluster (e.g. {\color{blue} Rajpurohit et al. submitted}), this confirms that polarised radio relics can also be formed in a tangled magnetic field. On the other hand, we showed that in a uniform ICM the polarisation fraction increases downstream, at odds with observations.
    
    \item[ii)] \textit{Alignment of the intrinsic polarisation $E$-vectors}: We found that a high degree of alignment at the shock front can also reproduced in a turbulent environment. 
    In particular, we find that the distribution of polarisation $E$-vector angles strongly depends on the upstream properties leading to differences between the two turbulent  models studied. We conclude that these differences between models are likely to be an outcome of the amount of amplification of $B_{\perp}$, which indirectly depends on the properties of the turbulent media such as the characteristic magnetic field scale $\lambda_B$ among others.
    This alignment, together with variations in the polarisation fraction in the downstream regions, 
    is observed too for example in the Sausage relic \citep[e.g.][]{2021arXiv210206631D}.
    
    \item[iv)] \textit{Internal Faraday rotation}: Our results suggest that at 6.5 GHz, the effect of Faraday depolarisation is marginal leaving the intrinsic trends unaffected (see point i). On the other hand, already at 1.5 GHz we find different degrees of Faraday depolarisation depending on the upstream turbulent configuration. Finally, Faraday depolarisation becomes significant at 150 MHz. Our 2L/3 model leads to a non-Gaussian RM distribution similarly to what has also been observed in cosmological simulations \citep[e.g.][]{wittor2019}. We caution that non-Gaussianity could result in discrepancies with respect to the common assumptions of symmetric RM distributions \citep[e.g.][]{1966MNRAS.133...67B,1991MNRAS.253..147T,1998MNRAS.299..189S}.
    
     \item[v)] \textit{Beam depolarisation}: Our study shows that Gaussian convolution for the limited beam size, strongly affects  the polarisation fraction and orientation of the polarisation $E$--vectors in a turbulent ICM. 
     In particular, we observe the effect of beam depolarisation when going to lower resolutions. For example, the mean polarisation fraction can drop down to $\sim10$--30 \% at a resolution of 12.5", depending on the turbulent model and the observing frequency. We find that the orientation of the polarisation $E$--vectors is mainly affected at the downstream of the shock. We find that the distribution of polarisation angles at a resolution of 12.5" tends to narrow down. As a consequence, the polarisation $E$-vectors may appear to be aligned even if the underlying medium is turbulent.
     
\end{itemize}

In summary, we have identified key
features that are likely to affect also the polarised emission from radio relics forming in a turbulent ICM. Our work confirms that a shocked turbulent ICM can reproduce the observed polarisation fraction from the shock surface, in contrast to a uniform ICM. We find that the degree of alignment of the $E$-vectors with the shock normal also depends on the level of underlying turbulence, and that a high alignment of the polarisation $E$-vectors at the very shock front, as observed in most of the radio relics, can be found in a turbulent medium. We expect higher Mach number shocks to produce higher polarisation fractions at the shock front. Yet, the observed polarisation in relics would be hard to reproduce with $\mathcal{M}\sim 2$ due to the low level of compression (see also \citetalias{dominguezfernandez2020morphology}). We compared our results to the mean polarisation fraction observed in some relics such as those in the galaxy clusters MACS J0717.5+3745, CIZAJ2242.8+5301, Abell 2744 and ZwCl\,0008.8+5215, and found that a turbulent ICM can reproduce comparable values in some cases. At the same time, our results highlight the complexity of reproducing one-to-one relations with observations, and of generalising the upstream characteristics to all relics. Indeed, there is substantial diversity of radio relics depending on the individual merger history of each galaxy cluster as well as on the viewing angle. In future work, we plan to tackle this issue by studying different viewing angles in combination with a broader range of initial turbulent ICM conditions considering different rms Mach numbers, magnetic field strengths, injection scales and types of forcing.

\section{Acknowledgements}

We thank the referee for useful comments that significantly improved the quality of this paper.
The analysis presented in this work made use of computational resources on the JUWELS cluster at the Juelich Supercomputing Centre (JSC), under
project "stressicm" with F.V. as principal investigator and P.D.F as co-principal investigator. The 3D visualization of the synchrotron emissivity was done with the VisIt software \citep[see][]{HPV:VisIt}.\\
P.D.F was partially supported by the National Research Foundation (NRF) of Korea through grants 2016R1A5A1013277 and 2020R1A2C2102800.
P.D.F, F.V., M.B. and K.B. acknowledge the financial support from the European Union's Horizon 2020 program under the ERC Starting Grant "MAGCOW", no. 714196. W.B.B. acknowledges financial support from the Deutsche Forschungsgemeinschaft (DFG) via grant BR2026/25. D.W. acknowledges financial support from the DFG via the grant number 441694982. \\
We thank S.O'Sullivan and F. de Gasperin for the useful comments and fruitful discussions.

\section{Data Availability Statement}

The data underlying this article will be shared on reasonable request to the corresponding author.

\bibliographystyle{mnras}
\bibliography{paola,franco}

\appendix

\section{Beam effects in the uniform medium}
\label{sec:app1}

\begin{figure*}
    \centering
    \includegraphics[width=0.27\textwidth]{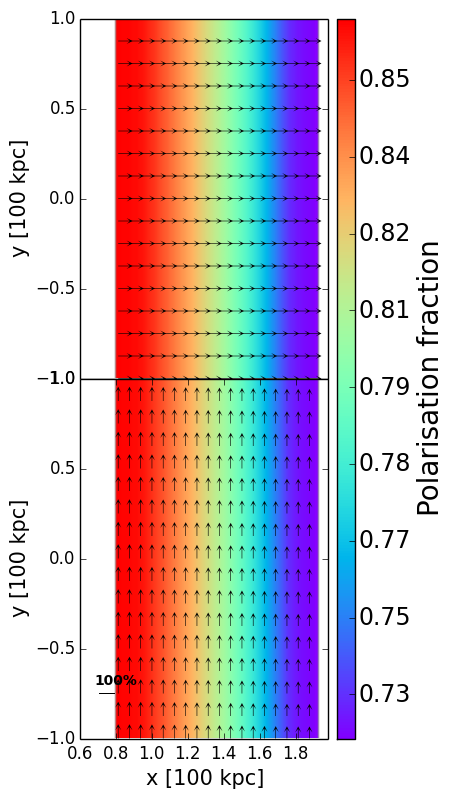}
    \includegraphics[width=0.27\textwidth]{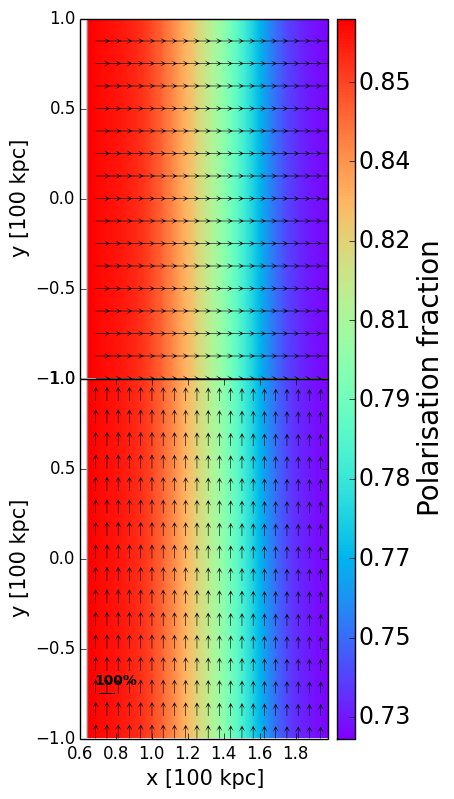}
    \includegraphics[width=0.27\textwidth]{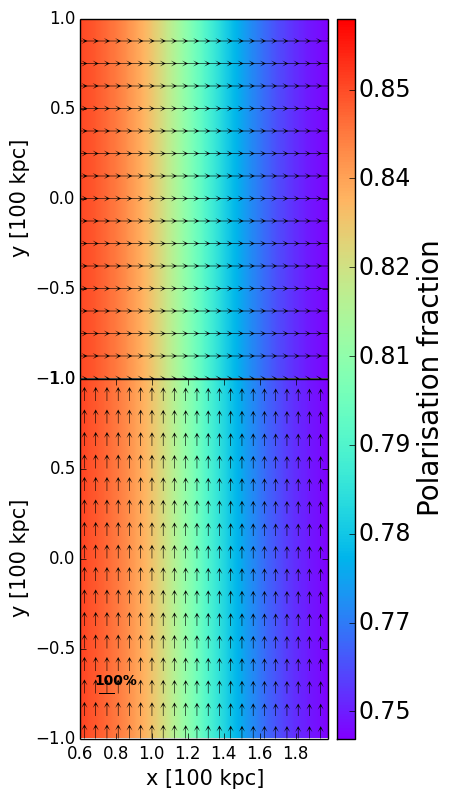}
    \caption{Smoothed polarisation fraction maps of the uniform medium considering $\theta_{\rm obs}=0^{\circ}$ at 150 MHz. We show the two magnetic field orientations: in the $y$--direction (\textit{upper panels}) and the $x$--direction (\textit{lower panel}). The first, second and third columns correspond to beam sizes of 2"$\times$2", 5"$\times$5" and 12.5"$\times$12.5" respectively. We overplot the polarisation $E$-vectors.}
    \label{fig:beam_uni}
\end{figure*}

In Fig.~\ref{fig:beam_uni} we show the maps of the polarisation fraction and the corresponding polarisation $E$-vectors, in a uniform media with different magnetic orientations (see Table \ref{table:init2}). This figure shows the same configuration as  Fig.~\ref{fig:pol_maps_UNI}, i.e. considering and observing angle of $\theta_{obs}=0^{\circ}$ and 150 MHz. The polarisation fraction maps are smoothed with a Gaussian kernel, considering the same three beam sizes as in Sec.~\ref{sec:beam}. We observe that the gradient of polarisation fraction, spanning from $\sim 0.7$ to $\sim 0.8$ in both magnetic field orientations, remains unchanged as in Fig.~\ref{fig:pol_maps_UNI}. Moreover,in a uniform medium the alignment of the polarisation $E$-vectors remains unchanged even at lower resolution. The main effect of taking into account different beam sizes in an uniform medium, is the apparent extension of the downstream region. In particular, the shock front region with a low polarisation fraction of $\sim 0.7$ and an extension of $\lesssim 20$ kpc (purple region) can appear to be twice as large for the 12.5"$\times$12.5" beam size. As we suggested in the analysis of the turbulent medium, and as expected, a resolution higher than 5"$\times$5" would more accurately reproduce the maps of polarisation fraction. 

\end{document}